# Nanogap-Engineered Core-Shell-Like Nanostructures for Comprehensive SERS Analysis

Mihai C. Suster†¶, Aleksandra Szymańska‡†¶, Tomasz J. Antosiewicz†, Agata Królikowska‡[*], and Piotr Wróbel†[*]

†Faculty of Physics, University of Warsaw, Pasteura 5, 02-093 Warsaw, Poland

‡Faculty of Chemistry, University of Warsaw, Pasteura 1, 02-093 Warsaw, Poland

¶Contributed equally to this work

E-mail: akrol@chem.uw.edu.pl; Piotr.Wrobel@fuw.edu.pl

## ABSTRACT

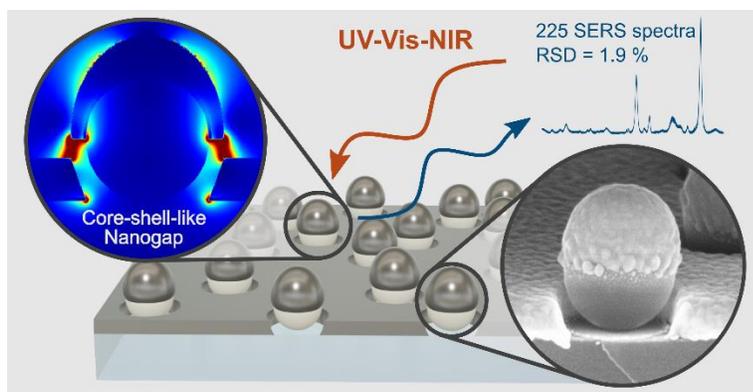

Development of fabrication protocols for large-area plasmonic nanostructures with sub-10 nm gaps with a spatially controlled distribution is critical for their real-world applications. In this work, we develop a simple, cleanroom-free protocol for the fabrication of macroscopic-sized plasmonic substrates (>6 cm$^2$), featuring a tunable multi-resonance optical response and light concentration in sub-10 nm gaps. Critically, these gaps are free to interact with the surrounding medium. This architecture consists of non-periodically distributed dielectric nanospheres coated with a metal multilayer, forming semi-spherical core-shell-like nanostructures (CSLNs) surrounded by a planar film. The sub-10 nm gaps formed between metal caps and the planar film are easily tuned by adjusting fabrication parameters such as multimetal layer thickness, composition, or nanosphere size and density. The excellent structural homogeneity, wide optical tunability, and extreme light confinement in the spatially controlled subwavelength nanogaps make CSLN-based substrates an ideal platform for comprehensive surface-enhanced Raman scattering (SERS) spectroscopy. This is proven through a combination of numerical modeling and iterative fabrication/characterization, leading to the optimized substrates showing cutting-edge spatial uniformity down to 1.9% determined as the relative standard deviation (RSD) of the SERS signal of p-mercaptobenzoic acid for 225 spectra over 3600 µm$^2$ area. High sensitivity is evidenced by an enhancement factor of ~10$^6$. The proposed substrates also meet all other demanding criteria, including sufficient signal temporal stability (RSD <4%), high substrate-to-substrate reproducibility (<15%), and SERS activity towards three various analytes. The unique geometry and wide spectral tunability of the CSLN substrates will also be of great value for other plasmon-driven applications.

# Introduction

Interaction of light with metallic nanogap structures can result in extraordinary trapping and amplification of the electromagnetic (EM) field in regions down to the sub-10 nm scale due to the excitation of localized surface plasmon resonances (LSPRs). Such strong focusing of the EM field allows for studying new phenomena related to light-matter interaction, including enhanced absorption,[1] nonlocality and ultra-strong coupling,[2, 3] quantum tunneling,[4] and charge transfer-plasmons.[5] Plasmonic nanogap field enhancement also enables development of more efficient energy and light sources, photodetectors,[3-5] and provides an ideal platform for ultrasensitive detection of molecules,[6] including surface-enhanced Raman scattering (SERS) spectroscopy[7, 8] with sensitivity reaching the single-molecule level.[9-11]

The most common approaches for fabricating plasmonic nanostructures featuring sub-10 nm nanogaps are based either on self-assembly of metal nanoparticles or nanolithographic techniques, like electron and focused ion beam lithography. An example of the former is nanosphere lithography (NSL), which allows relatively simple and large-area fabrication of nanogap structures. In NSL, dielectric nanospheres (DNSs) are assembled over a support to form a uniform, hexagonally packed monolayer. Using this approach, along with a combination of ion etching and atomic layer deposition techniques, several architectures have been developed, featuring nanogaps in the few-nm regime. Examples include a metal film over nanosphere (MFON) structure,[7, 12-14] self-assembled nanorings with sub-10 nm gaps,[15] and nanoforests with gold nanoparticles deposited on top of a hexagonal array of nanopillars.[16] In the latter case, a metal thickness-dependent nanogap is formed between the nanoparticle and the discontinuous gold film deposited in the interpillar voids on the substrate. However, the aforementioned approaches either suffer from a lack of repeatability and insufficient resolution of the nanogap size – not reaching below 10 nm, or they produce structures featuring plasmonic hot spots inaccessible to the external environment and exhibiting a non-specific broad optical response. Due to these limitations, there is still a growing demand for a simple, low-cost, and high-throughput technology capable of fabricating large-area plasmonic nanostructures with easy-to-access gap sizes below 10 nm, offering a rich and tunable optical response.

The development of new nanofabrication procedures is especially desired in the field of SERS spectroscopy, which is an outstanding analytical method capable of identifying and quantifying chemicals with single-molecule sensitivity.[17-19] This sensitivity is achieved by enhancing the otherwise very weak Raman scattering signal. The efficiency of SERS relies on two distinct mechanisms: electromagnetic field enhancement and chemical effect (mainly due to charge transfer resonances).[9, 20, 21] Strongly dominant electromagnetic contribution to the total amplification of the Raman signal is possible by the use of noble metal plasmonic nanostructures, which is why in recent years they have become somewhat synonymous to SERS. Mainly due to its high sensitivity and flexibility of design, SERS is widely used in a variety of areas including analytical, imaging, environmental, pharmaceutical, medical, chemical, biological, and security-related applications.[22-28]

Although SERS effect has been recognized for over 50 years as an excellent scientific tool, reproducibility of the results remains a challenge.[29] Indeed, this handicap is known to the research community to be a major limitation to the widespread use of SERS outside the laboratory, what triggered the establishment of rigorous standards for evaluating SERS-active plasmonic substrates.[30, 31] Specifically, five main criteria for a high-quality SERS-active substrate were formulated:[30] (1) high spatial reproducibility defined as less than 20% spot-to-spot variation of the SERS signal over 10 mm$^2$; (2) high substrate-to-substrate reproducibility, validated by less than 20% SERS signal variation over 10 substrates of the same type; (3) sufficient temporal stability allowing for a maximum 20% SERS signal variation measured weekly for a month; (4) high sensitivity with the recommended enhancement factor exceeding 10$^5$ over an area of at least 7500 μm$^2$; and (5) documented SERS activity toward three analytes not exhibiting surface-enhanced resonance Raman scattering (SERRS) established for neutral, positively and negatively charged adsorbate molecules. Despite many efforts dedicated to successful fulfillment of these demanding standards, the task remains unresolved.[9] Critically, many works emphasize optimization of the substrates in just one of the aforementioned areas without considering the others.

Herein, we develop a simple, scalable, and easily tailored method for fabricating remarkably spatially uniform plasmonic substrates of macroscopic size (>6 cm$^2$) that meet all five criteria described above as vital for comprehensive SERS analysis. These substrates consist of metal-coated dielectric nanospheres with easily adjustable sub-10 nm gaps formed in a single evaporation process. The proposed nanoarchitecture supports a core-shell-like geometry formed by simple metalization of a solid support decorated with non-periodically distributed dielectric nanospheres. Each individual nanostructure features an easily accessible nanogap and exhibits multiple tunable plasmonic resonances, both of which are particularly important for sensing applications. The energy of the symmetric (nanogap), antisymmetric, and quadrupole plasmonic modes can be both coarse- and fine-tuned by adjusting one or more of several available fabrication parameters, including the size and concentration of dielectric nanoparticles or the material composition and thickness of the metallic coating.

Most importantly, control of the thickness of the evaporated material results in the formation of a nanogap of desired size in the sub-10 nm range with single-nanometer precision, leading to extreme light confinement and corresponding EM field enhancement. The measured absorbance spectra of the substrates are corroborated by finite-difference time-domain (FDTD) simulations, thus enabling prediction of fabrication parameters resulting in a specific set of excitations at desired resonance wavelengths. Such nanogap-engineered plasmonic substrates exhibit efficient and well-defined light-matter interactions, demonstrating not only a strong and temporally stable enhancement in SERS spectroscopy but also an extremely spatially uniform surface enhancement over an area exceeding 6 cm². This development also paves the way for plasmonic platforms with customizable nanoarchitectures based on tunable core-shell-like nanostructures, suitable for various applications that require both a tunable, multimodal optical response and equally important point-to-point and sample-to-sample reproducibility.

## Results and discussion

**Fabrication of Core-Shell-Like Nanostructures (CSLNs)**

The key steps of the proposed protocol of fabricating Core-Shell-Like Nanostructures (CSLNs) are schematically shown in Figure 1. The fabrication process is based on the nanosphere lithography framework.[32-34] First, negatively-charged sulfate latex Dielectric NanoSpheres (DNSs) are deposited on a solid glass support covered initially with a positively charged thin layer of poly(diallyldimethylammonium chloride) (PDDA) (see Figure 1a and b). Electrostatic forces between same- and opposite sign electric charges assure surface trapping of particles and their separation,[33] as well as a short-range correlated (amorphous) distribution of the DNSs.[35] Next, directional physical vapor deposition (PVD) of a germanium wetting layer and a thin plasmonic metal film (Figure 1c and d) creates a cap over each nanosphere separated from the planar metal layer deposited between the DNSs. This results in the formation of CSLNs with uniform nanogaps between the caps and the planar film (see Figure 1d for the schematic representation and Figure 1f for the actual geometry imaged using scanning electron microscopy (SEM)). The size of these nanogaps can be easily controlled by adjusting the evaporated film thickness and the diameter of the DNSs. This approach, while being a modification compared to conventional NanoSphere Lithography (NSL) methods, remains straightforward, and allows to achieve a unique geometry with features in the sub-10 nm regime.

As presented in Figure 1e, the SEM image providing a top view of a representative area of the whole substrate produced in this manner clearly demonstrates that the proposed experimental protocol results in a homogeneous coverage with an amorphous array of nanospheres with no sign of aggregation. Uncontrolled aggregation is a common concern in bottom-up approaches of nanostructure assembly onto a solid support, which translates into an inhomogeneity of the resulting geometrical arrangement. Particularly, in Surface-Enhanced Raman Spectroscopy (SERS) analysis, this leads to spectral irreproducibility related to random formation of highly localized regions of intense EM fields, so-called "hot spots". In our approach, suppression of oligomerization is ensured by strong nanosphere-PDDA polymer electrostatic attraction and mutual repulsion between the same-sign charged latex nanospheres, which jointly overcome interparticle van der Waals and Casimir attraction (see SEM images in Figure S1 in the Supporting Information (SI) and the discussion in Section S1 therein).

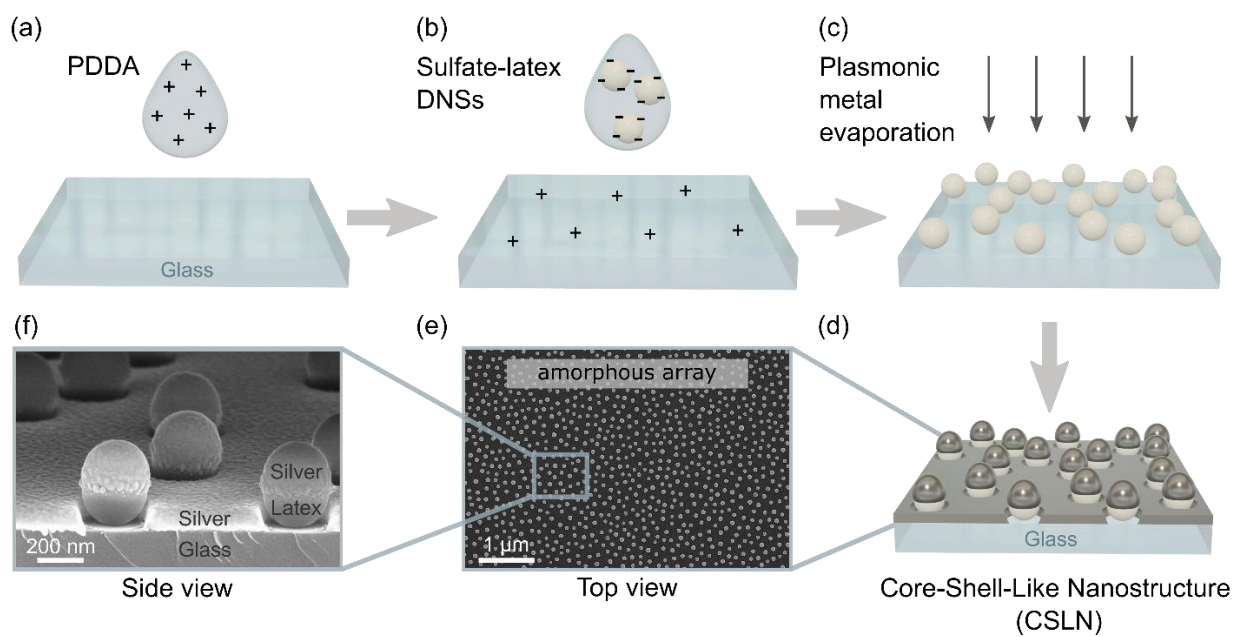

**Figure 1. Schematic illustration of the fabrication process of an amorphous array of CSLNs: (a) electrostatic attraction between a layer of positively charged PDDA polymer initially deposited on a glass support and (b) following drop-cast negatively charged sulfate-latex DNSs results in an amorphous array of DNSs. (c) The array is then covered by a plasmonic metal using PVD technique, (d) leading to the formation of the CSLN geometry featuring nanogaps between the metal caps and the planar metal layer formed on the glass support. (e) Large-scale SEM image (top view) of a global distribution of CSLNs within an amorphous array and (f) SEM image (side view) showing individual CSLNs with recognizable metal caps and nanogaps.**

The SEM image in Figure 1f shows a side view of a typical nanostructure obtained herein. The upper half of the DNS is covered with a metal cap, whose maximum thickness at the very top of the sphere is equal to the nominal thickness of the planar layer on the substrate. The thickness on the side of the DNS decreases on average to less than 30% of the nominally evaporated value (see Figure S2 to examine the details). Such a geometry is ascribed to an interplay of two coexisting effects. The first one is related to different wettability and varying metal atom flux normal to the local particle surface, resulting from the curvature of the DNS. The second effect is due to the shadow cast by the nanosphere, which determines the diameter of the aperture within the planar metal film deposited on the glass around the DNS (see SEM images in Figure S2 for details). This shaded area, inaccessible to metal atoms during evaporation, increases with the thickness of the metal layer at the side of the sphere. Hence, the aperture walls tilt away from the DNS, as can be seen in Figure 1f.

Continuity of both the cap and the bottom metal layer is achieved by evaporating a 1.5 nm thick Ge film, which serves as a wetting layer to prevent island-growth of the silver.[36] This subtle modification of the fabrication protocol leads to major changes in the resulting nanostructure's geometry: in the absence of Ge, the wettability of DNSs is very low and therefore the evaporated silver forms one ill-defined bulk nanoparticle on top of a single DNS. The substrate uniformity in this case is very low as each DNS is covered with a metal nanoparticle of different shape and size. An increase in the wettability allows silver to create a well-adhering cap around the upper half of each DNS (compare respective SEM images and absorbance spectra in Figure S3 in SI and the discussion therein, section S1) with the final structure resembling a core-shell geometry. Germanium also smooths out the planar metal layer, resulting in a well-defined toroidally-shaped nanogap between the aperture edge and the cap (see SEM images in Figure S2). Increasing the thickness of the metal film naturally leads to closing of the nanogap, which formally is expected at the evaporated thickness corresponding to the radius of the DNS. However, due to the directionality of the evaporation process and the aforementioned shadow effect, the aperture size increases upon metal layer growth, and thus the gap is still present for metal film thicknesses exceeding the DNS radius size.

**FDTD calculations**

FDTD calculations were performed to predict and explain the optical response of the CSLNs. Figure 2a shows the absorbance spectrum calculated for a nanostructure with DNSs of 60 nm in diameter

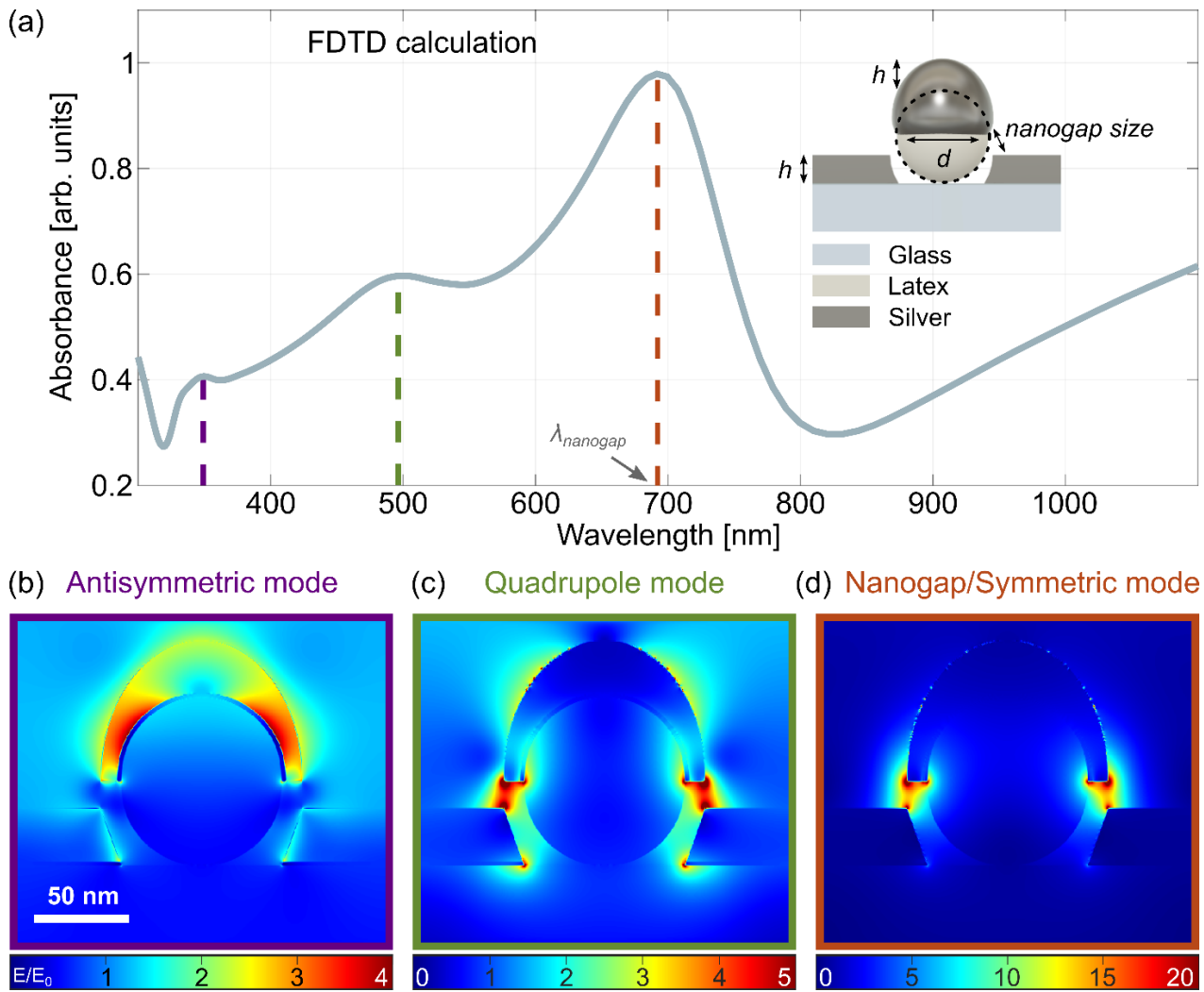

**Figure 2.** (a) FDTD-calculated absorbance spectrum of CSLN with 60 nm dielectric core diameter (*d*), and Ag layer thickness (*h*) of 20 nm, resulting in a nanogap size of 8.5 nm (see the model geometry on the right). (b)-(d) Electric field distributions for the structure depicted in (a) corresponding to the plasmonic modes calculated at the resonant wavelengths indicated by the color-coded dashed lines in the plot shown in (a). The types of the plasmonic modes are given in the legend.

(*d*), 1.5 nm thick Ge layer, and Ag nominal thickness (*h*) of 20 nm, which results in a gap size of 8.5 nm (*cf.* the scheme in the inset of Figure 2a for the model of a single simulated nanostructure). The thickness of the cap at the side of the DNS is 30% of that on the top, according to the experimental data (*cf.* SEM images and parameters in Figure S2). The simulated absorbance spectrum exhibits three distinctive plasmon resonances associated with the electric field distributions depicted in Figure 2b-d, where the colors of the frames match those of the dashed lines in Figure 2a, indicating the respective spectral positions. The most intense maximum at ~700 nm (marked as $\lambda_{nanogap}$ in Figure 2a) corresponds to a symmetric nanogap mode formed between the metal cap and the edge of the aperture within the planar metal layer (Figure 2d). The field amplitude associated with the resonance at ~500 nm exhibits a higher-order, quasi-quadrupolar nature (Figure 2c). However, it shows notable deviations from the classical quadrupole mode observed for the bulk metal nanospheres due to coupling of the two bottom lobes with the nanogap mode. The enhanced EM field surrounds the nanogap, ensuring the largest amplitude is located therein. The electric field distribution for the resonance peak at ~385 nm exhibits a maximum at the DNS/metal cap interface (Figure 2b) and strongly resembles the antisymmetric mode characteristic for the core-shell nanoparticles.

The calculations indicate that the CSLNs exhibit a response reminiscent of core-shell plasmonic particles, although modified by coupling to the aperture. The response of a typical core-shell plasmonic nanosphere stems from coupling of two dipolar modes with one being that of a metal sphere and the other of a void in metal. This interaction yields a symmetric dipole mode across the whole particle and a higher order antisymmetric dipole resonance, whose spectral separation depends on the metal shell thickness. As the shell/core ratio increases (at a constant core size), the spectral separation of the two resonances diminishes, with the higher order mode decreasing in amplitude and the remaining single resonance acquiring the characteristic properties of a dipole mode of a bulk metal nanoparticle.[37] By observing such a trend for CSLNs, one could confirm that the metal-capped DNSs located in apertures exhibit similar behavior to fully-coated core-shell nanoparticles. Therefore, the influence of the metal layer thickness on the optical response of the CSLN structure was calculated and confronted with the experimental data. Figure 3 compares the absorbance of CSLNs based on a 60 nm DNS coated with a silver layer with thickness varying from 5 nm to 25 nm as measured in the experiment (Figure 3a) and calculated *via* the FDTD method (Figure 3b). Both sets of curves show a similar characteristic trend. Increasing the thickness of the silver layer leads to a blue shift of the energies of both the nanogap and quadrupole modes, while the resonance in the UV range is shifted towards longer wavelengths. Apart from the overall increase of the absorbance due to the presence of a larger amount of lossy material, the amplitude of the plasmonic resonances is substantially improved with increasing silver thickness, which is likely related to diminishing nonlocal effects and reduced electron scattering in thicker metal films.

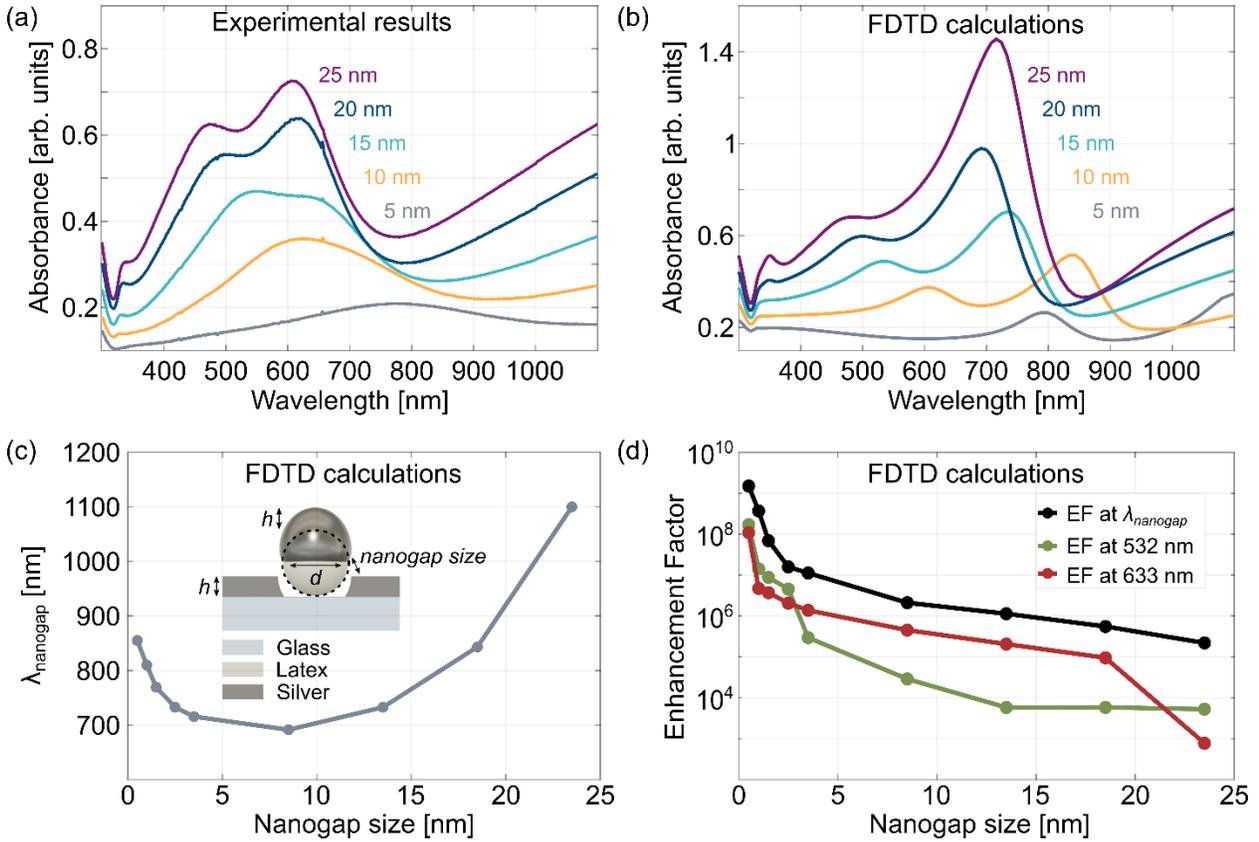

**Figure 3.** The absorbance spectra of CSLNs based on DNSs with 60 nm dielectric core diameter (*d*) and Ag layer thickness (*h*) varying from 5 nm to 25 nm: (a) as measured in the experiment, and (b) as calculated *via* the FDTD method; (c) FDTD-calculated relationship between the wavelength of the nanogap resonance and silver layer thickness (calculated for nine values of *h* and *d* = 60 nm for model geometry in the inset); (d) semi-log plot of FDTD-calculated values of enhancement factor (*EF*) for varying gap size at the nanogap resonance wavelength); and at Raman excitation wavelengths: 532 nm and 633 nm (calculated in each case for nine values of nanogap size and *d* = 60 nm); The size of the nanogap for panels (c) and (d) is expressed in simulations as $d/2 - h - h_{Ge}$, where $h_{Ge}$ is the thickness of Ge (1.5 nm), while the lines connecting the points are a guide to the eye.

Overall, the optical response of the analyzed CSLN is consistent with that of a conventional core-shell nanosphere, whose FDTD calculated electromagnetic properties are presented in Figure S4 (extinction, scattering and absorption cross-sections, and electric field distribution for 60 nm diameter dielectric core and 20 nm Ag shell) and Figure S5 (extinction cross-section as a function of the silver layer thickness for 60 nm DNS), further discussed in the section S2 of the SI.

The main features of the experimental absorption data and the effect of the thickness of the plasmonic metal film are well reproduced by FDTD simulations (*cf.* Figure 3a and b). Discrepancies between experimental and numerical data in terms of the exact positions of resonances and their width and amplitudes are typical differences between the idealized model and the experimental conditions. It is worth stressing that the simulations do not take into account any imperfections observed in the experiment, such as dispersion of the nanosphere size, surface roughness, as well as the tendency of silver to form granular features at the rim of the cap rather than a solid and well-defined edge. Moreover, due to the directionality of the evaporation process, substrates placed off the rotation axis may exhibit slight asymmetry of the metal cap leading to modified spectral response of the CSLNs. Finally, the amorphous and thus disordered nature of the nanoparticle array means that while the average optical response of such an array may be fairly well reproduced by simulating the properties of a single particle, either alone or using subwavelength periodic boundary conditions, some quantitative shifts of the exact position, amplitude, and width of the resonances are expected.[38] Additionally, absorbance measurements were performed with an incoherent light source which decreases the resonance contrast in comparison to simulations assuming a coherent linearly polarized plane wave.

Figure 3c shows the FDTD-calculated changes of the spectral position of the plasmon resonance corresponding to the mode due to nanogaps of varying size for 60 nm DNSs. For the 8.5-23.5 nm range of the nanogap size, this resonance mode exhibits a blue shift as the size of the nanogap decreases (and thus the thickness of the silver film increases), which is typical of the symmetric mode in core-shell nanoparticles. In contrast, for nanogap size less than 8.5 nm we observe a red shift and further increase in the discussed resonance peak amplitude with decreasing gap size (increasing metal thickness), as can be also seen in simulated absorbance spectra in Figure 3b. The origin of the resonance red shift is a strong coupling between the plasmons excited at the cap and those at the aperture edge, which is a typical optical response for the nanogap structures.[39, 40] In other words, as the gap size decreases, two different behaviors of this resonance peak occur, confirming its dual origin predicted by FDTD. For gaps much larger than 10 nm, the resonance undergoes a blue shift like in a typical core-shell nanostructure, and for gaps smaller than 10 nm it red shifts like for a standard nanogap mode (as seen in Figures 3b and 3c). This leads also to a large electromagnetic field enhancement that can be expressed as an enhancement factor ($EF$), calculated as $EF \sim |E/E_0|^4$, where $E$ and $E_0$ are amplitudes of the electric field of the resonant plasmon mode (here: nanogap mode) and the incident light beam, respectively; assuming that the incident and inelastically scattered photons are of similar energy. The semi-log plot in Figure 3d shows the numerically calculated $EF$s as a function of the nanogap size for the CSLNs architecture at $\lambda_{nanogap}$ (nanogap resonance wavelength), as well as 532 nm and 633 nm wavelengths, used as excitation lines in the following SERS experiments. For nanogap sizes above 5 nm, the $EF$ is in the range of $10^6$, while for smaller gaps it increases to $10^7$ and reaches $10^9$ for the gap sizes below 1 nm (Figure 3d). Use of photons with wavelengths within the plasmon resonance peak, but not matching exactly its maximum, results in a decrease of $EF$ by only one order of magnitude, which means that $EF$ of around $10^8$ can be reachable in practice.

## Optical response and tuning

Small changes of the easily adjustable parameters of the CSLN fabrication protocol can result in drastic or minimal modifications in the plasmonic properties of the final structure, depending on the modified parameter. This makes the CSLN a very convenient and practical platform as its optical features can be easily tailored specifically to meet the desired requirements of a particular application. The process of tuning the spectral position of one or more of the available plasmon resonances can be divided into two stages: coarse and fine spectral tuning. Coarse tuning involves modifying the fabrication parameters that induce major spectral changes by shifting the resonance wavelengths by many tens or even hundreds of nanometers. Then, the position of the resonances can be fine-tuned with resolution reaching single nanometers by adjusting other, much less impactful parameters that only slightly change the resonant energy while preserving the overall optical response of the entire photonic system.

There are about 10 relevant parameters which can be adjusted during the CSLN fabrication process to affect the absorbance spectra, enabling shifts in the energy of a specific plasmonic mode of the substrates. Representative examples of spectral tuning using selected single experimental factors, along with corresponding SEM images showing the resulting structural differences in CSLNs, are presented in Figure 4. First, coarse spectral tuning is achieved *e.g.* by using DNSs of different diameters (Figure 4a, top row). This change also affects the overall topography of CSLNs in terms of the particle density (*cf.* SEM images for 60 nm and 300 nm DNSs in Figure 4a, bottom row) and the nanogap size and its optical response for a given Ag thickness. For example, changing the diameter of the DNSs from 100 to 200 nm (*i.e.*, by 100%) spectrally shifts the gap mode by approximately 250 nm (from 640 to 890 nm), as can be seen in Figure 4a, top row. The two other plasmonic modes, *i.e.*, the quadrupole and antisymmetric modes, are also red-shifted with an increasing diameter of the DNSs, but to a much lesser extent.

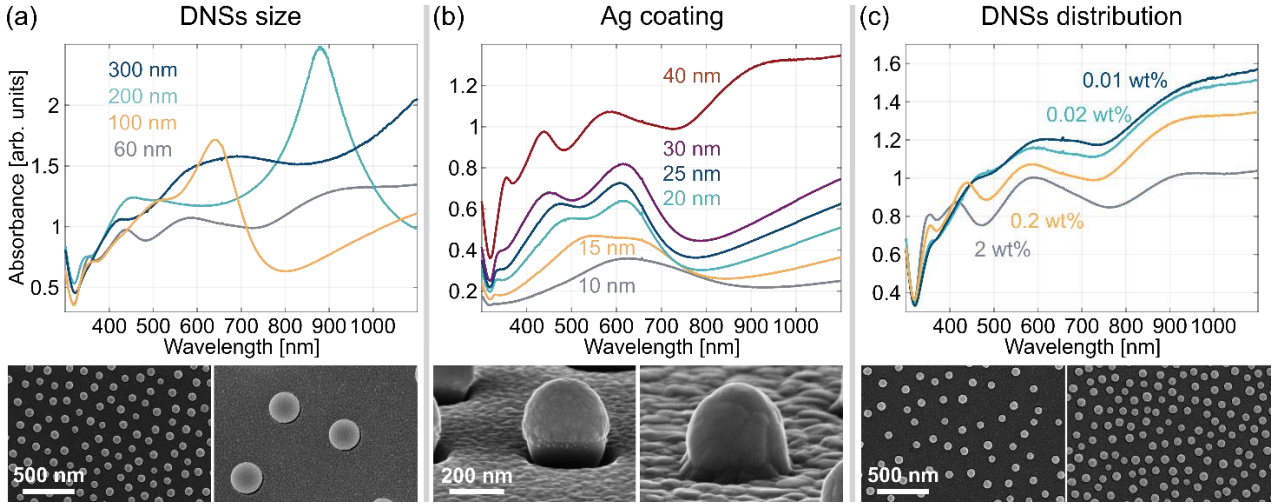

**Figure 4. Demonstration of rough and fine spectral tuning capabilities within the CSLN architecture: (a) rough tuning by using DNSs of different diameters (fabrication parameters: 0.2 wt% suspension of 60-300 nm DNSs, metal multilayer: 1.5 nm Ge and 40 nm Ag); (b) fine-tuning by varying the silver layer thickness (0.2 wt% suspension of 60 nm DNSs, 1.5 nm Ge, 10-40 nm Ag); (c) ultrafine tuning by changing the mean distance between the DNSs, controlled by their bulk concentration (0.01-2 wt% suspension of 60 nm DNSs, 1.5 nm Ge, 40 nm Ag). The values of (a) the DNS diameter, (b) the nominal thickness of the silver layer and (c) the bulk concentration of DNSs suspension are given in the legend. Top row: absorbance spectra showing the spectral tuning characteristics of a given fabrication parameter. Bottom row: SEM images illustrating pronounced effects of a given parameter on the morphology of the CSLN.**

Next, fine-tuning can be realized, *e.g.*, by varying the thickness of the metal coating. Figure 4b (top row) illustrates that as the silver layer thickness increases, the quadrupole peak mode undergoes a blue shift. For example, evaporating 30 nm of Ag instead of 15 nm (*i.e.*, changing $h$ by 100%) shifts the quadrupole mode by *ca.* 70 nm (from 510 to 440 nm). An analogous blue shift for an increasing metal shell thickness accompanied by a simultaneous narrowing of the peak width is observed for the nanogap plasmonic mode. However, one has to keep in mind that an additional effect of an increasing metal thickness is a gradual closing of the nanogap (see SEM images for a 300 nm diameter DNS coated with 100 and 300 nm of Ag shown in Figure 4b, bottom row). Finally, the antisymmetric mode shifts consistently toward the red as the Ag layer is growing, and its intensity increases significantly when the nominal shell thickness exceeds half the DNS diameter (*cf.* absorption spectra for $h$ = 30 and 40 nm for 60 nm DNS in Figure 4b, top row).

Finally, ultrafine tuning can be accomplished *e.g.* by heavily diluting the suspension of DNSs, which results in an increased mean interparticle distance (see SEM images for 2 wt% and 0.01 wt% for 60 nm DNSs in Figure 4c, bottom row). This allows for subtle resonance peak shifting (Figure 4c, top row) due to modification of interparticle coupling.[38] All three observed resonance peaks are red-shifted as the average distance between the DNSs increases with the decreased concentration of the suspension, though not necessarily with equal magnitude nor monotonously across the whole examined parameter range, which is caused by different phase delays between each individual mode.[41] The total substrate coverage by DNS decreases with lowering of the bulk concentration, therefore the contrast of all the peaks deteriorates as the overall structure acquires a reflective character with fewer DNSs dispersed on the substrate. The antisymmetric mode is the least affected, since a change of DNSs concentration from 0.02 to 0.01 wt% red-shifts the quadrupole mode by less than 10 nm. Hence, DNS concentration is a very convenient parameter for ultrafine tuning of the optical response because a significant, and thus easy to precisely control, dilution causes subtle (even single nanometer) changes in the position of resonances.

Using a combination of only these three parameters already enables generation and positioning of plasmonic resonances across the whole visible (Vis) and near-infrared (NIR) spectral range. If needed, further tuning can be performed, *e.g.*, by deposition of the CSLNs on a different support, using DNSs with a changed refractive index, or evaporating a more sophisticated multilayer consisting of a number of materials (metals and/or dielectrics), and many more. The main reason for choosing these three parameters in the tuning procedure described above is that they can be easily adjusted in a continuous manner. The DNSs can be synthesized in any size in the range from around 40 nm up to a few μm, the plasmonic metal film can be evaporated with a sub-nm thickness precision, and the interparticle distance can be controlled by heavily diluting the DNSs suspension. This provides an elegant and straightforward strategy for tailoring the plasmonic properties of CSLNs by perfectly matching a target wavelength imposed by a specific application. Moreover, the presence of three plasmonic resonances makes these substrates particularly useful for SERS spectroscopy, since it is possible to tune the peaks simultaneously (as well as individually) to the energies of the excitation line and the Stokes bands of the molecules, thus enabling efficient, double or even multi resonance SERS measurements. [42, 43]

Having established the physical justification of the optical modes of the CSLN nanostructure which yield the field enhancement necessary for SERS effect, we consider one final aspect. While silver is the superior metal for large field enhancements in visible light excited plasmonic applications, it does not offer good chemical stability under all conditions. Hence, for practical applications it is recommended to consider passivation layers which will protect silver against corrosion.[44] However, use of additional dielectric layers on top of a plasmonic metal will substantially lower the field intensity experienced by molecules, lowering the *EF* significantly. Hence, gold, due to its greater chemical inertness and biocompatibility, is a better choice than silver in (bio)sensing, despite the larger losses from interband transitions extending far into the visible.[45] Thus, we leverage the best parameters of both materials by using the low-loss silver as a bottom layer, and coating it with 5 nm of Au. This should ensure good optical properties while significantly enhancing stability in the experiment and protecting the nanostructure from the harmful effects of harsh solvents. As plotted in Figure S6, an additional 5 nm Au layer on top of 20 nm of Ag for 60 nm DNSs predictably shifts the nanogap resonance to the position similar to 25 nm of Ag. This shows that the overall evaporated metal thickness, and thus the resulting nanogap size, governs the position of the nanogap resonance. On the other hand, the addition of 5 nm Au leads to a diminished quadrupole mode due to the interband transition in gold occuring for wavelengths below 516 nm.[46] Thus, bearing in mind the minor effect of gold on the essential features of surface plasmon resonances (see Figure S6 and discussion in S3 of the SI), we employ only CSLN nanostructures with a variable amount of silver which is capped by 5 nm of Au in the SERS characterization that follows.

## SERS performance of CSLNs substrates

We now turn to the examination of SERS performance of the proposed CSLNs plasmonic platform with respect to the five criteria that determine a high-quality SERS substrate, as specified in the Introduction section. Validation of these demanding requirements is carried out *via* comprehensive SERS studies, which confirm the fulfillment of all five conditions by the developed substrate. For a succinct summary of the outcomes of this study we refer to Table 1, while below we introduce the experimental details followed by a presentation and discussion of the results.

SERS validation of the CSLN substrates is performed using p-mercaptobenzoic acid (pMBA; see inset in Figure S7 for molecular formula) which is also known as 4-mercaptobenzoic acid (4-MBA) and can be considered as a representative of a neutral and/or anionic molecule, depending on the conditions of the experiment. The SERS signature of pMBA is well-recognized since it was first reported by Michota and Bukowska.[47] The monolayer of pMBA is formed through self-assembly, where the molecules are firmly attached to the surface by covalent bonding between the metal (both Ag and Au) and sulfur. In addition, pMBA exhibits a large Raman scattering cross-section typical of benzene derivatives and its SERS signal is sensitive to the pH of its surroundings.[48, 49] All these features established pMBA as the most commonly used Raman reporter in pH nanosensors,[50-52] as well as a labeling molecule in SERS nanotags for bioimaging.[53, 54]

Typically all SERS experiments were carried out for self-assembled monolayers (SAMs) of pMBA grown overnight on CSLN substrates to ensure the formation of a defect-free and complete monolayer, unless the volume concentration of the thiol in solution was insufficient (as discussed further for the pMBA concentration dependent SERS studies). In all cases, the substrates incorporate a dual metal Ag/Au layer, where the thickness of the former is a variable parameter and the latter is always 5 nm thick to provide protection from harsh chemicals. A typical SERS spectrum excited with a 532 nm laser observed for pMBA adsorbed from a $10^{-4}$ M ethanolic solution on CSLNs arrays is presented in Figure S7, while the vibrational assignment of the most intense SERS bands can be found in the inset in Figure S7.[47, 50, 55, 56] The appearance of vibration bands characteristic for carbonyl group stretching (around 1700 cm$^{-1}$) and symmetric stretching of the carboxylate anion (around 1360 cm$^{-1}$) indicates partial deprotonation of pMBA upon adsorption. The two bands that dominate the SERS spectrum of pMBA are due to benzene ring breathing modes: $\nu_{12}$ at 1075 cm$^{-1}$ and $\nu_{8a}$ at 1585 cm$^{-1}$.[22] Their intensities are extremely useful to quantify the SERS signal of pMBA not only due to their large surface enhancement, but also because of the independence of their intensity on environmental parameters (except those affecting molecular orientation). In the conducted evaluation of CSLNs' performance, the main focus is on the 1585 cm$^{-1}$ band as it is the strongest one in the SERS spectrum of pMBA.

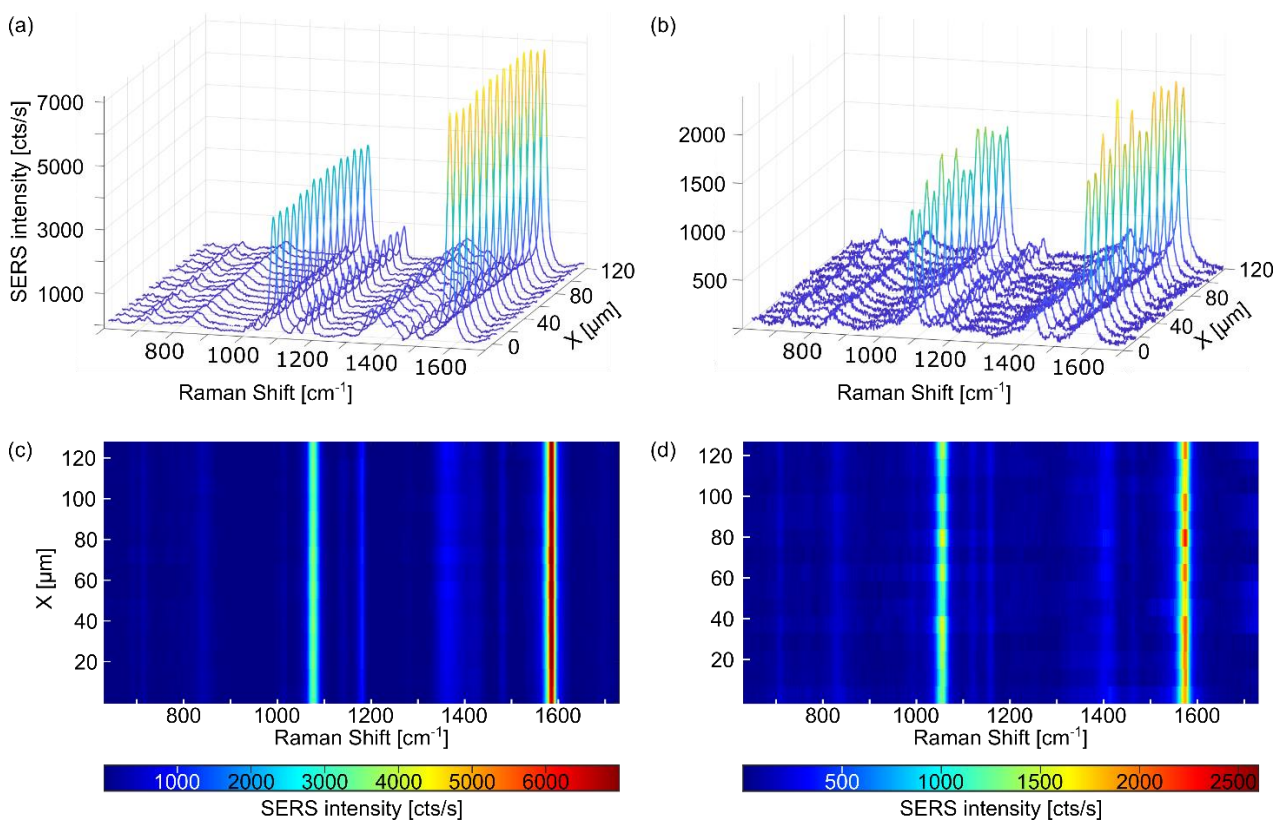

**Figure 5. Data representative of excellent and mediocre SERS performance, illustrating the spectral fluctuations by 3-axis graph visualization and corresponding waterfall plots of SERS spectra of 10$^{-4}$ M pMBA collected using mapping mode from 15 different spots (line profile) of the CSLNs substrates fabricated with 0.2 wt% suspension of DNS, coated with 1.5 nm Ge, 40 nm Ag and 5 nm Au for: (a) and (c) 60 nm DNS and 532 nm excitation line *vs* (b) and (d) 200 nm DNS and 633 nm excitation line.**

Prior to demonstrating a systematic procedure for optimizing the fabrication protocol of CSLNs as substrates for SERS spectroscopy, we present an explicit illustration of how synthesis and measurement conditions can affect the SERS response of the substrates. Figure 5 compares the two sets of SERS spectra acquired for two CSLN substrates that differ in fabrication parameters and Raman data collection conditions, selected to illustrate excellent *vs* significantly worse quality in terms of their SERS performance. The top panel presents line profiles through the SERS map captured point-by-point and depicting the intensity distribution over the entire Raman shift range studied for pMBA adsorbed on CSLNs substrates fabricated from 0.2 wt% DNS suspensions and coated with 1.5 nm Ge, 40 nm Ag and 5 nm Au, but differing in DNS size and SERS spectrum excitation line: 60 nm and 532 nm (Figure 5a) *vs* 200 nm and 633 nm (Figure 5b), respectively. For the green laser and 60 nm DNSs, only minor fluctuations in the intensity of the SERS spectra can be observed, and the signal itself is very strong. For the red excitation line and 200 nm DNSs, one can see much more pronounced differences in the intensity of the SERS signal, which is also less enhanced, as visible in Figure 5b. The advantage of the waterfall plots presentation is that in addition to the differences in the intensity of the SERS spectra, one can better see the fluctuations in vibrational energy, which practically do not occur for 60 nm DNS and the 532 nm laser (Figure 5c), while their strong contribution for 200 nm DNS and the 633 nm laser (Figure 5d) suggests the inhomogeneous chemical state of the pMBA monolayer under measurement conditions. These results confirm the need to optimize the fabrication procedure of CSLNs as a key step for the comprehensive SERS performance and to successfully meet all five criteria that define a high-quality SERS substrate.

We begin by characterizing the impact of DNS size (Figure 6, left column), thickness of the Ag layer deposited below the 5 nm of Au (Figure 6, middle column), and bulk concentration of DNSs in suspension (Figure 6, right column) on the SERS signal intensity of pMBA at 1585 cm$^{-1}$. Absorbance spectra of the fabricated CSLNs are shown in Figure 6a-c. The average SERS signal and corresponding standard deviation (SD) acquired with two excitation wavelengths (532.0 nm and 632.8 nm) for the three analyzed fabrication parameters are shown in Figure 6d-f, while the RSD values of these results are plotted in Figure 6g-i. For the statistical needs of the SERS analysis, the signal was collected in each single case from a total of 675 various locations across three $60 \times 60$ μm$^2$ areas on the substrate separated by at least 1 cm, each covering 225 distinct spots.

The dependence of SERS intensity of the 1585 cm$^{-1}$ band with varying DNS size is similar for both excitation lines (see Figure 6d). The strongest signal is observed for diameters 60 and 100 nm, comparable for each laser and in both cases it decreases with a particle size. We recall here that the optimal conditions for SERS occur when both the incident and Raman scattered photons are plasmon-enhanced. For single resonance substrates the plasmon peak should be located halfway between the wavelengths of the excitation laser and the Raman-shifted band of the analyte while covering them both.[57-59] In the case of a multiply-resonant substrate, it is preferred to match energy of one plasmonic peak to the Rayleigh band and the other to the Stokes bands.[60, 61] Figure 6a shows a red shift of the nanogap resonance with an increase in DNS diameter, which significantly changes the matching of this intense mode with the excitation wavelengths (the solid green/red vertical lines representing the 532 and 633 nm excitation lasers, respectively) and the corresponding absolute values of Stokes-shifted band (position in maximum of the SERS band) of pMBA at 1585 cm$^{-1}$ (dashed vertical lines at 581 nm and 703 nm). The strong SERS signal observed in Figure 6d for 60 and 100 nm nanospheres is a result of a very good overlap with the nanogap mode for both excitation lines, although the 633 nm laser is slightly red-detuned and thus provides a weaker SERS signal. This contrasts the significantly smaller SERS signal for the 200 and 300 nm DNS substrates whose nanogap modes are shifted to 830 nm and 1100 nm, respectively. In these two cases there is a significant mismatch between the excitation/Stokes energies and the plasmon mode, resulting in weaker surface enhancement.

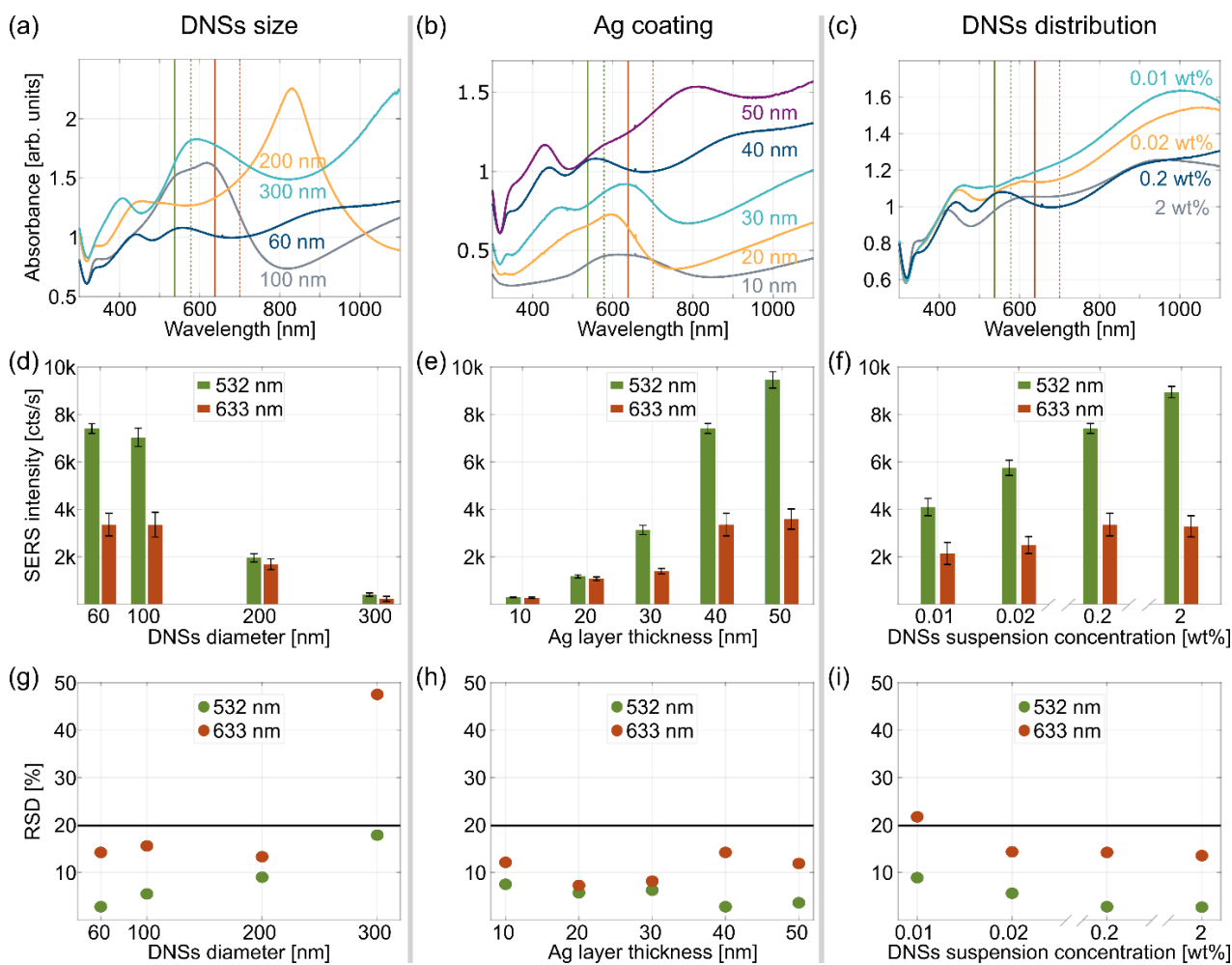

**Figure 6.** Absorbance curves analogous to the CSLN substrates analyzed in a top row of Figure 4, but with an additional 5 nm thick gold layer on top presenting the effects of different: (a) diameter of DNS (fabrication parameters: 0.2 wt% suspension of 60-300 nm DNSs, metal multilayer: 1.5 nm Ge, 40 nm Ag, 5 nm Au); (b) silver layer thickness (0.2 wt% suspension of 60 nm DNSs, 1.5 nm Ge, 10-50 nm Ag, 5 nm Au); (c) mean distance between the DNSs (0.01-2 wt% suspension of 60 nm DNSs, 1.5 nm Ge, 40 nm Ag, 5 nm Au). The values of (a) the DNS diameter, (b) the nominal thickness of the silver layer and (c) the concentration of DNSs suspension are given in the legend. Effect of varying fabrication parameters of CSLNs analyzed in the top row on (d) - (f) SERS intensity of pMBA band at 1585 cm$^{-1}$ and (g) - (i) on the corresponding RSD values of SERS intensity. Each bar or point represents averaged data from 675 SERS spectra measured over an area of about 140 mm$^2$. The solid black line at 20% indicates the maximum acceptable level of RSD in SERS analysis, as defined in Table 1. Green and red coding represents the results for a particular excitation wavelength (see the legend). The solid and dashed lines in a top row indicate the wavelengths corresponding to excitation lasers and Stokes-shifted analyzed SERS bands, respectively (see the text for details).

Interestingly, the decrease of the SERS intensity is much larger for the 532 nm laser: approximately 3.6-fold versus 2-fold for the 633 nm line when going from a diameter of 100 nm to 200 nm. Moreover, an observed magnitude of the SERS signal becomes nearly identical for these two excitation wavelengths. To elucidate the latter, we recall that the second factor affecting SERS intensity is the nanogap size and its impact on the overall EM field enhancement. Both the quadrupole and symmetric plasmonic modes are characterized by an enhanced electric field in the nanogap (see Figure 2c and d), however, the value of this enhancement decreases rapidly with an increasing gap size (Figure 3d). Thus, it is likely that this decrease is an additional, important reason for the sharp drop in SERS intensity for $d > 100$ nm. This is because the nanogap size increases with the diameter of the DNS for a constant metal thickness, as is illustrated by the SEM images in Figure S8 (the nanogap sizes are *ca.* 5, 20, 50, and 90 nm for the four analyzed DNS diameters). Based on the $|E/E_0|^4$ values calculated using FDTD results (Figure 2c and d), we estimate that the electric field enhancement is more than one order of magnitude larger for the nanogap/symmetric resonance than for the quadrupole mode. Thus, matching the excitation/Stokes band energy with the quadrupole mode yields weaker SERS enhancement when compared to the symmetric nanogap mode. These factors result in a similar SERS signal for both laser lines for the 200 nm DNS, because both the excitation lines and Stokes bands do not coincide with any plasmon resonance and exhibit only weak enhancement due to a gap of intermediate size (19 nm according to Figure S8b). For $d = 300$ nm the gap size becomes larger and the overall enhancement decreases even more. However, the overlap of the quadrupole mode by both excitation lines is comparable and rather mediocre, while there is a better match with Stokes-shifted band for a green laser, which explains why the SERS signal excited with 532 nm wavelength is stronger than for the 633 nm. These results confirm how important the presence and size of nanogap are in the context of optimal SERS performance. For large nanogaps, regardless of the position of its plasmonic resonance with respect to the laser lines Raman bands, the enhancement will not be satisfactory for visible light excitation. An additional factor to consider is the surface density of DNSs and thus the number of the nanogaps acting as EM hot spots available to the molecules. A simple approximation confirmed by SEM images in Figure 4a and surface density scaling of the random sequential adsorption indicates that for bigger DNS particles their surface density and thus the number of nanogap SERS-active hot spots decrease, which also contributes to a general reduction of SERS intensity for larger DNS sizes.

Figure 6g presents the RSDs corresponding to the SERS intensities discussed above and plotted in Figure 6d. An exceptionally low RSD value of 2.5% is found for the 60 nm DNS excited using 532 nm, while for the larger 100 nm particles it remains excellent at 5.5%. While RSD increases with nanosphere size, it does not exceed the required 20% limit (marked with a black solid line in Figure 6g and following ones in the same row) for the green laser excitation. Higher RSD values are noted for all DNS sizes in combination with the red laser, however, only for the $d = 300$ nm case they exceed the 20% criterion (#1 in Table 1). The latter can be attributed to the fact that for diameters larger than 200 nm, the probability of DNS oligomerization increases drastically.

Two factors contribute to the rather general increase of RSD with nanosphere diameter. Firstly, for a constant metal thickness, it is most likely related to the increasing nanogap size, as a small nanogap is characterized by a very uniform distribution of the EM field. Thus, it is expected that the smallest gap observed for the 60 nm DNS (Figure S8a) exhibits simultaneously the smallest deviations and largest values of the SERS signal, confirming that ultrasmall nanogaps are essential for reproducible and strong SERS signal.[62] A secondary reason is related to the variability of the size of metal-coated nanospheres and their distribution on the surface. The SD of the overall diameter increases for the larger DNSs sizes (see Table S1 for the data determined from SEM images for the metal-evaporated nanospheres). Since the size of the nanospheres determines its resonance wavelength, a broader size distribution will increase the spread of the plasmonic peaks contributing jointly to the total optical response of a given area on the substrate and thus to the properties of SERS signal. Indeed, the distribution of the size of CSLNs (Table S1 and relevant short discussion in the section S4 of SI) perfectly matches the trends observed for the RSD of the SERS signal in Figure 6g. Furthermore, the size of the nanosphere determines their surface number density, meaning that for a given laser spot size the number of particles over which the SERS signal is averaged decreases with an increasing DNS diameter. As the arrays are random with a short range correlation, the immediate vicinity of each particle is unique. Thus, in such an amorphous array even consisting of identical in size particles, the individual resonances will vary due to different radiative coupling.[41,63] Averaging SERS signal over fewer nanospheres when using a constant laser spot size means that the variability of the smaller plasmon-active area of the substrate will be larger, further supporting the DNS-size-dependent RSD observations for both laser lines (Figure 6g).

Considering the factors discussed above, we ascribe the extraordinary spatial uniformity of the strong SERS signal observed for the smallest investigated DNS size and constant metal thickness mainly to the unique and statistically homogenous geometry of CSLNs and their relatively dense packing. This combines the benefits of size-controlled core-shell-like nanoparticles with small nanogap structures and adequate averaging over a number of plasmonic features (see Figure 1e and 1f, together with Figure S1 and S8a) resulting in a well-defined optical response (Figure 4a and Figure 6a).

Due to the very promising combination of both high surface enhancement and extremely low RSD values of the SERS signal for 60 nm DNSs, we proceed with this diameter and investigate the effect of Ag layer thickness on the SERS response of pMBA. Silver layers with thicknesses of 10, 20, 30, 40 and 50 nm along with 5 nm of gold on top were evaporated (absorbance spectra in Figure 6b, and the resulting SERS intensities and corresponding RSD values are shown in Figure 6e and 6h, respectively). The behavior of the plasmonic peaks is similar to that of only the Ge and Ag coating (*cf*. Figure 4b and the related discussion), although the total metal thickness in this case is larger, as it also includes 5 nm of Au. For the same total amount of metal only a slight resonance shift is noticeable, as depicted in Figure S6.

In Figure 6e we observe consistent behavior for both excitation wavelengths: the SERS intensity of the pMBA band at 1585 cm$^{-1}$ gradually increases with a growing thickness of the Ag layer. The metal thickness increase is directly responsible for a decrease of the size of the nanogap, which is the dominant factor affecting the SERS signal intensity. For each substrate, except for the 50 nm Ag-coated one, a strong nanogap resonance overlapping the 532 nm (laser) and/or 581 nm (Stokes) wavelengths is observed (Figure 6b). However, the substrate with the thickest examined Ag layer (50 nm) exhibits a complicated absorbance curve resulting from the combined optical response of CSLN statistically containing fully or partially closed nanogaps for which none of the major resonances are overlapped with both laser lines or Stokes bands. For the 633 nm line laser and associated 703 nm Stokes band, the resonance maximum of the symmetric nanogap mode does not perfectly overlap any of these two wavelengths for the majority of examined thicknesses of Ag layer, hence the SERS signal is typically weaker than under green laser excitation. This detuning increases for thicker Ag layers, especially for 50 nm of Ag, but the significant decrease of the nanogap size compensates for the spectral mismatch, still providing decent SERS intensity.

The RSD dependence of the SERS signal on Ag thickness exhibits a more complex pattern (Figure 6h) than it was in the case for the DNS size effect. For the 532 nm excitation line, we observe RSD values in the 5-10% range for the 10 - 30 nm Ag layer, and even a lower one for 40 nm Ag with a very small, but well-controlled, *ca.* 5 nm gap (*cf*. Figure S8a). Further increase of the metal thickness by an additional 10 nm begins to close the nanogap in a stochastic way due to partial growth of nanoislands around the inside rim of the nanoaperture, which means that the size of the nanogap can no longer be fully controlled. The ultrasmall sub-5 nm gap still provides efficient hot spots for SERS, but becomes inhomogeneous within the spot size. As plotted in Figure 3d, the *EF* is strongly variable on the minute differences in size within the sub-5 nm nanogap range. Thus, even small variability of the gap yields very large changes of the *EF*. Therefore, when changing *h* of Ag from 40 to 50 nm, we observe a simultaneous increase in RSD and SERS signal for the green laser line.

The pattern of RSD dependence for both lasers is the same, except for the largest two values of Ag layer thickness, when the RSD is markedly different. The main cause behind this discrepancy is likely a different behavior of field enhancement at small gaps for the green and red laser (*cf.* FDTD simulations in Figure 3d for the gap size ≤ 4 nm ) and a smaller overlap of the red excitation line and the corresponding Stokes band with the plasmon peaks. The resulting weaker overall plasmon response of the whole nanostructure leads to a comparatively larger contribution from the local inhomogeneities of the metal coating, which create additional isolated, single-nm hot spots that ultimately increase the RSD of SERS signal.

In general, the best conditions for large and uniform SERS enhancement are provided by operating near the gap closing, occuring in a uniform way around the circumference, while avoiding the formation of isolated sub-1 nm hot spots (see again the undesired in SERS experiment strong *EF* variations in Figure 3d for the gap size ≤ 4 nm). These conditions are best fulfilled for 60 nm DNS coated with 1.5 nm of Ge, 40 nm of Ag and 5 nm of Au, showing RSD of the SERS signal of 2.5 %. However, all considered metal thicknesses offer outstanding performance in terms of SERS signal homogeneity with RSD <7.5% for 532 nm laser and below 15% for 633 nm laser, thus easily within a tolerance range defined by criterion #1 in Table 1.

Subsequently, we optimize the surface density of the DNS by varying their suspension concentration when creating the templates for metal deposition. The results of this procedure, namely SERS intensities of the pMBA band at 1585 cm$^{-1}$ for varying bulk concentrations of DNS, are plotted in Figure 6f. The volume concentration of the DNSs determines the minimum center-to-center distance between particles deposited on PDDA coated glass support. With the increase of the DNS concentration, the number density of illuminated elements in the CSLN also increases, due to effectively decreased mean interparticle distance between deposited DNSs (*cf.* SEM images in the bottom row of Figure 4c). The distance between the nanospheres determines the relative phase of the incident and scattered light impinging onto the nanostructures, and by changing this relation all resonances are shifted spectrally, either to the blue or red – depending on the ratio between the resonance wavelength and the minimum center-to-center distance. The peak shifts are in the range of up to 10% of the single particle response, thus playing a minor, though non-negligible role, as the width of some of the plasmon peaks is rather broad. The more important parameter affecting SERS performance of the substrates is an increase of the particle number surface density with an increasing

DNSs concentration. It leads to a spectral response collected from a larger number of illuminated (laser-activated) CSLNs, which simultaneously increases the SERS signal (Figure 6f) and, due to better averaging over a larger number of DNSs, decreases the RSD (Figure 6i). The reason for the decrease in RSD values is similar to that observed for a decreasing size of the DNS particle (Figure 6d and g). The concentration-dependent RSD values are markedly smaller for green laser excitation (below 10%) than for the red one (13-22%). This excitation laser wavelength dependence is also present for the substrates with Ag layer (and 5 nm of Au) thicker than 30 nm, as visible in Figure 6h. As discussed previously, this difference is caused already by the geometry of the gap of an individual CSLN, thus a decreasing particle density is unlikely to significantly alter the RSD ratio between the green/red laser. The lowest and nearly identical RSD values of 2.7% and 2.8% occur for the concentrations of 2.0 wt% and 0.2 wt%, respectively (Figure 6i). Note that if we analyze 225 SERS spectra from an area of about 3600 $\mu m^2$ for the highest studied concentration of DNS, the RSD reaches an extremely low value of 1.9%. Despite the slightly better SERS signal for 2.0 wt% concentration, for further evaluation we select the CSLN substrates fabricated with 0.2 wt% suspension due to a significantly smaller amount of material used in their preparation, making it a more cost-effective option with respect to the end cost of the substrate fabrication. Guided by the preceding analysis, all further characterization of SERS performance utilizes CSLN substrates composed of 60 nm DNSs deposited from a suspension of 0.2 wt% and coated sequentially with 1.5 nm Ge, 40 nm Ag and 5 nm Au.

The next considered quality of the CSLNs is the enhancement factor. It is a quantitative parameter which allows comparison of various substrates in terms of their SERS performance.[61] Unfortunately, methods used for calculating the *EF* are often highly questionable. Common problems include neglecting the resonance Raman contribution to the overall *EF* and using unjustified, rough estimations of experimental parameters, with molecular surface coverage being the most difficult one to determine accurately.[64] Indeed, recent works have highlighted the lack of standard procedures in determining *EF* values, both in terms of measurement conditions and analyte selection.[29, 65, 66]

In order to minimize all these pitfalls, we use pMBA and pyridine (Pyr) to evaluate SERS activity of the CSLN substrates, mainly because these molecules do not exhibit resonance Raman cross-sections under experimental conditions. These analytes were also chosen because they are among the most recognized adsorbates by the SERS community – pMBA as a routine Raman reporter for pH sensing, and Pyr as the first molecule for which SERS effect was experimentally observed.[67] Moreover, we adopt an approach in which the *EF* is determined with reference to a standard electrochemically roughened silver substrate produced by an oxidation-reduction cycling (ORC) procedure. An additional benefit is that the presented results involve SERS measurements carried out using identical experimental parameters (including exactly the same ORC roughening procedure and Raman set-up) as the reference data from Ambroziak et. al.[68] In this work, *EF* for 0.05 M Pyr in 0.1 M KCl adsorbed on a reference ORC-roughened silver electrode was found to be equal to $4.6 \times 10^5$, as determined for the Pyr SERS breathing mode at 1004 cm$^{-1}$ ($\nu_1$, Wilson notation).[68,69] Here applied calculations involve comparing intensity of the SERS signal acquired under specific measurement conditions to that obtained under identical conditions for Ag ORCs with known surface enhancement parameters and determining the *EF* for CSLN substrates on that basis. This procedure provides *EF* values for CSLNs up to $1.5 \times 10^6$ when using only Ge and Ag, and $0.62 \times 10^6$ with an additional 5 nm Au layer. Such estimated *EF* values for Pyr for both excitation lines are summarized in Table 2. These parameters, determined from measurements over an area twice larger than the minimum recommended size, indicate excellent sensitivity of CSLN substrates for SERS spectroscopy (criterion #4 in Table 1). Moreover, these experimental results are in reasonable agreement with the theoretical *EF* values predicted for the nanogap mode (see Figure 3d), which are on the order of $0.3 \times 10^6$ for the green laser when the gap size is just below 5 nm (as determined from SEM image in Figure S8a). This confirms the dominating role of the nanogap plasmon mode in the SERS response of CSLNs.

The *EF* for pMBA cannot be estimated using the above method due to a lack of appropriate reference data. Therefore, the so-called SERS gain ($G_{SERS}$) is used instead for this analyte (see Table 2 for the determined values). It is calculated as a ratio of the intensity of the SERS signal for the CSLN substrates and the Raman signal for the bare and 5 nm Au coated 40 nm planar silver layer (normalized to laser power and acquisition time). This approach allows for directly establishing the impact of the CSLN geometry, with material composition aside on SERS performance. $G_{SERS}$ is a less common indicator of SERS substrate sensitivity than *EF*, but it enables a straightforward evaluation of the signal amplification by a given nanostructure relative to a planar metallic film[70,71] or a normal Raman signal for a volume sample of the same concentration.[72] Advantageously, it is not affected by an uncertainty related to the estimation of the number of molecules probed in SERS and the reference measurement. SERS signal for CSLNs excited with 532 nm laser for the 1585 cm$^{-1}$ band of pMBA is more than two orders of magnitude higher than for the reference planar layers, achieving $G_{SERS}$ of 177 for 40 nm Ag and 113 when using the additional 5 nm of Au. This difference is even more pronounced for the red laser, for which $G_{SERS}$ is an order of magnitude larger for the substrate without 5 nm Au than in its presence, among others as a result of almost 4 times stronger SERS signal for CSLNs for the former. While these values are valid for the CSLN substrates considered optimal in terms of their comprehensive performance, it is apparent that $G_{SERS}$ for pMBA is much smaller than the actual *EF* value. This is due to the significantly smaller number of molecules experiencing the enhanced EM field of the hot spots in the nanogaps relative to the reference planar substrates measurement.

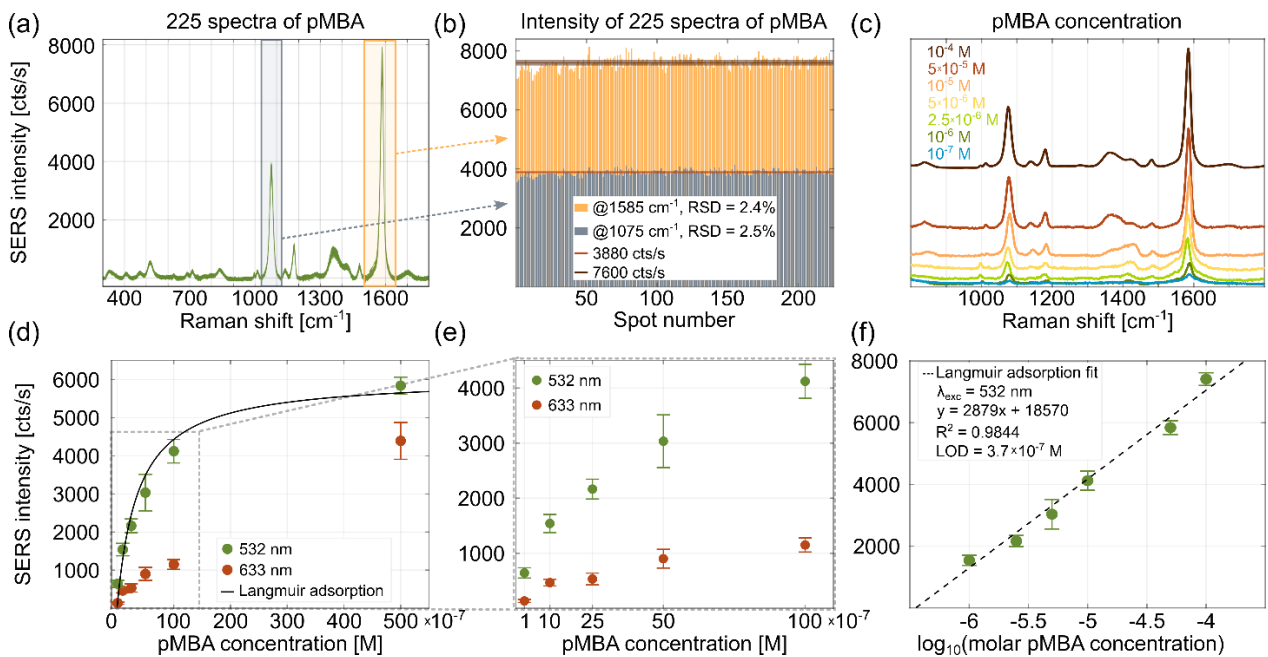

**Figure 7. (a) SERS spectra of $10^{-4}$ M pMBA adsorbed on an amorphous array of CSLNs, collected from 225 substrate spots covering an area of around 3600 μm$^2$. (b) The intensity variation of two characteristic pMBA bands obtained from the data in (a) showing an RSD value of ≤2.5 %. (c) SERS spectra excited with a 532 nm laser for varying molar concentrations of pMBA from $10^{-7}$ to $10^{-4}$ M. (d) SERS intensity of the 1585 cm$^{-1}$ band *vs* pMBA molar concentration for the two excitation lines and fitted Langmuir adsorption model for the green laser; and (e) its zoomed-in region up to $10^{-5}$ M. (f) Linear relationship between the SERS intensity of pMBA band and the logarithm of analyte concentration (parameters of the fit and LOD for pMBA given in the legend). All data were acquired for optimized CSLNs, fabricated with 0.2 wt% suspension of 60 nm DNS, coated with 1.5 nm Ge, 40 nm Ag and 5 nm Au. SERS data in (a)-(c) and (f) were acquired with a 532 nm laser.**

In Figure 7a, one can see the excellent repeatability of the 225 SERS spectra of pMBA collected from a square-shaped area of about 3600 μm$^2$ (excited with a 532 nm laser). This is illustrated further in Figure 7b with a bar plot of SERS intensities for said spectra acquired from 225 different substrate spots for pMBA. The obtained RSD values are 2.4% and 2.5% for bands at 1075 cm$^{-1}$ and 1585 cm$^{-1}$, respectively. Increasing the analyzed substrate area to 140 mm$^2$ and the number of SERS spectra to 675 nm only slightly increases the RSD value to 2.8%, far exceeding the requirements for a high-quality SERS substrate (criterion #1 in Table 1).

This extraordinarily low RSD value of 2.8% obtained from a large substrate area (>100 mm$^2$) and a large number of points (nearly 700) is a cutting edge result when considering the spatial uniformity of SERS-active substrates. One of the few examples of such exceptionally low RSD of the SERS signal for a comparable number of points are the very recently reported substrates based on ultrastrong coupling between the LSPR of gold nanoparticles and Fabry-Pérot nanocavities, which provide RSD of 3% for 625 locations within a 30 × 30 μm$^2$ area.[73] Similar RSD values of less than 3%, although for an undefined number of SERS measurements, were also demonstrated for substrates with nanogaps based on a different geometry than proposed here, however, the structures were produced with sophisticated deep UV lithography.[74]

The performance of optimized CSLNs in SERS-based quantitative analysis of pMBA is used to further characterize the fabricated substrate. In Figure 7c we plot SERS spectra obtained for pMBA concentration values ranging from $10^{-7}$ M to $10^{-4}$ M excited with 532 nm wavelength. The changes of SERS intensity with a varying pMBA concentration are analyzed using the signal of the 1585 cm$^{-1}$ band for the two examined laser lines. As seen in Figure 7d and the magnification of low concentrations range in Figure 7e, the SERS signal initially increases rapidly with rising pMBA concentration and reaches a quasi-plateau when the concentration is $\geq 10^{-5}$ M. This saturation of the SERS signal is typical of surface adsorption[75] and is consistent with the Langmuir adsorption isotherm model (see the fit in Figure 7d for the green laser), for which the surface coverage of the analyte is given by:

$$\frac{Q}{Q_m} = \frac{KC}{1+KC}, \tag{1}$$

where $Q/Q_m$ is the surface coverage given by a ratio of the SERS intensity of the analyzed band for a particular concentration of pMBA molecules in solution ($C$) and when all adsorption sites are occupied, while $K$ is the adsorption equilibrium constant. The observed trend suggests that for pMBA concentrations lower than $10^{-5}$ M, thiol-metal interactions are partially hindered with only a portion of the active sites being occupied by pMBA molecules. A probable cause of this behavior is the progressive dissociation of the carboxyl groups, which is supported by the presence of the symmetric stretching vibration band of $COO^-$ groups in the SERS spectra (see Figure 7a and the band assignment in Figure S7), thus evidencing a partial deprotonation of pMBA already for the $10^{-4}$ M concentration. The carboxylate form of pMBA may preferentially interact with the metal and/or electrostatically repel molecules diffusing out of solution, both of which will hinder adsorption and lower the resulting surface concentration of the thiol (even reaching a submonolayer regime). Next, the concentration dependent SERS data for the 532 nm excitation line and the 1585 cm$^{-1}$ band are used to establish the limit of detection (LOD) of pMBA. A linear fit to the semilogarithmic plot of the SERS-signal-*vs*-pMBA-concentration data is performed and plotted in Figure 7f. Using a protocol involving the analysis of a blank sample (see section Methods for details), we obtain the LOD as $3.7 \times 10^{-7}$ M. It is worth highlighting that the LOD for pMBA obtained with the optimized CSLNs is one order of magnitude lower (and thus better) than that for commercial gold SERS substrates, while the RSD is also reduced by one order of magnitude.[76] This demonstrates the improved sensitivity and precision of the SERS measurement with the substrates developed in this work.

The shelf life of a SERS substrate is also a critical parameter for real-world applications and mass production. Therefore, we compare the SERS signal of optimized CSLN substrates soon after fabrication and after 4 months of storage before functionalization with $10^{-4}$ M of pMBA. This analysis is performed for substrates with and without an additional 5 nm Au layer and the results are presented in Figure S9a and b, respectively. While changes of the SERS signal were not monitored weekly, the 4 month storage of the substrates significantly exceeds the recommended period of one month for testing temporal stability (criterion #3 in Table 1). The drop of the SERS signal of pMBA at the 1585 cm$^{-1}$ band is 18.7% for the CSLNs covered with Ge/Ag/Au multilayer and 3.6% for the Ge/Ag one. For the 1075 cm$^{-1}$ band the decrease of SERS intensity does not depend significantly on the presence of the Au layer and is equal to 13.6% with the Au and 15.8% without it. The smaller decrease in intensity for the pMBA SERS signal of the at 1585 cm$^{-1}$ band in the absence of Au is surprising. However, due to the increased complexity of the multilayered system, it is possible for material migration between layers to occur,[77] which would result in a modification of plasmonic characteristics and thus the observed change in SERS intensity. A second possibility is a subtle change of the orientation of the pMBA molecular monolayer due to surface restructuring by segregation within the metallic layer, the former of which can result in a change in the relative intensity ratio of the SERS bands.[75]

The previously made decision to apply a 5 nm Au film coating is further justified by the significantly lower and competitive RSD values of SERS signal compared to using Ag alone. The lower RSD value with gold capping is confirmed for both freshly prepared CSLN substrates (for a 532 nm excitation and 1585 cm$^{-1}$ band the RSD is 2.8% with Au *vs* 11.3% without) and 4 months after fabrication (6.2% *vs* 8.5%, respectively). Additionally, Au-coated substrates are substantially more resistant to aggressive reagents; for example, they remain stable when exposed to chloride ions in solution.

Substrate-to–substrate reproducibility of CSLNs was also verified by collecting SERS response of pMBA for 6 substrates fabricated over a 5-month timeframe. The evaluation is based on SERS measurements from a total of more than 20 locations on these 6 substrates (see Figure S10), further cut into smaller pieces. Each spot corresponds to 225 spectra from a 60 μm x 60 μm squared area. The variability of the SERS signal on this sample of more than 4500 spectra from different fabrication batches is more than satisfactory, as evidenced by RSD of 14.7%. We therefore consider that CSLNs meet the condition for high quality SERS substrates (criterion #2 in Table 1).

In the final step of the evaluation of CSLN SERS performance, we use Pyr and methyl orange (MO) under off-resonance conditions to examine SERS capability toward detection of analytes under different charge states, *i.e.* neutrally/positively/negatively charged adsorbates. Specifically, Pyr can undergo partial protonation in aqueous solution even at neutral pH, while MO is a halochromic anionic/neutral azo dye which contains a pH sensitive chromophore. As can be seen in Figure S11, a reliable and consistent SERS signal (*cf.* the RSD values of the SERS intensity given in the legend) is readily obtained for these two analytes next to pMBA for the optimized CSLN substrates, demonstrating versatility of this platform (criterion #5 in Table 1).

**Table 1.** Verification if CSLNs satisfy the criteria of a high-quality SERS substrate.[30, 31] The assessment involves the SERS band of pMBA at 1585 cm$^{-1}$ for optimized substrates fabricated with 0.02 wt% suspension of 60 nm DNSs, 1.5 nm Ge, 40 nm Ag and 5 nm Au, unless stated otherwise.

| Criterion type | Required | Observed for CSLNs | Evaluation |
|---|---|---|---|
| #1 High spatial reproducibility | Less than 20% spot-to-spot variation of SERS signal over 10 mm$^2$ | Top result: RSD of 2.7% for the most spatially uniform substrates (2 wt%) and RSD of 2.8% for the optimized substrates, measured at 675 points over approximately 140 mm$^2$ (see Figure 6g, 7a, and 7b) using a 532 nm laser. RSD of 1.9% for 225 spectra and *ca.* 3600 μm$^2$ area (for 2 wt%). | Cutting-edge result |
| | | Less than 20% for all the substrates except the largest DNS size (300 nm) and the lowest DNS concentration (0.01 wt%) for 632.8 nm laser (see Figure 6g-i) | More than satisfied |
| #2 High substrate-to-substrate reproducibility | Less than 20% SERS signal variation over 10 substrates of the same type | RSD of 14.7% for 21 spots (225 spectra from *ca.* 3600 μm$^2$ rectangular area for each spot) distributed across various fragments originating from macroscopically cut sections of 6 optimized substrates of the same type fabricated within 5 months (see Figure S10) | Satisfied |

| #3 Sufficient temporal stability | 20% or less SERS signal variation, measured weekly for a month | Deviation from original SERS intensity of 3.6% for the optimized substrate without Au and 18.7% for the optimized substrate with Au, measured once after 4 months from fabrication (see Figure S9) | More than satisfied |
|---|---|---|---|
| #4 High sensitivity | Values of $EF$ exceeding $10^5$ over an area of at least 7500 μm$^2$ | $EF$ typically above $0.33 \times 10^6$ for optimized substrate with Au and up to $1.5 \times 10^6$ for optimal substrate without Au, determined over an area of approximately 17 000 μm$^2$ (see Table 2) | Satisfied |
| #5 SERS activity toward three analytes | SERS signal for three chemicals not exhibiting SERRS effect: with positively charged, neutral and negatively charged adsorbate molecules (for each case) | SERS activity for the optimal substrates documented towards pMBA (anionic/neutral), Pyr (cationic/neutral), and MO (methyl orange; anionic/neutral) under non-resonant conditions (see Figure S11) | Satisfied |

Table 2. Determined values of an enhancement factor ($EF$) for Pyr and SERS gain ($G_{SERS}$) for pMBA adsorbed on CSLNs fabricated with 0.2 wt% 60 nm DNSs, 1.5 nm Ge, 40 nm Ag with and without 5 nm Au.

| Analyte; concentration and SERS band | Parameter | 40 nm Ag | 40 nm Ag, 5 nm Au |
|---|---|---|---|
| *532 nm laser* | | | |
| Pyr; $5 \times 10^{-2}$ M in 0.1 M KCl@1010 cm$^{-1}$ | $EF$ | $1.5 \times 10^6$ | $0.62 \times 10^6$ |
| pMBA; $1 \times 10^{-4}$ M,@1585 cm$^{-1}$ | $G_{SERS}$ | 177 | 113 |
| *633 nm laser* | | | |
| Pyr; $5 \times 10^{-2}$ M in 0.1 M KCl @1010 cm$^{-1}$ | $EF$ | $0.46 \times 10^5$ | $0.33 \times 10^6$ |
| pMBA; $1 \times 10^{-4}$ M, @1585 cm$^{-1}$ | $G_{SERS}$ | 1486 | 150 |

## Conclusions

Nanogap plasmonic structures garner considerable attention in various fields of science and engineering due to large electromagnetic field enhancement and subwavelength mode volumes of excited surface plasmon resonances. However, their real-world applicability is limited by the lack of convenient and cost-effective nanofabrication methods capable of spatial control of matter in the sub-10 nm regime on macroscopic-size substrates. To address this issue, we presented a simple and clean-room-free procedure for fabricating plasmonic substrates of a few cm$^2$ in size, featuring nanogaps of engineered dimensions, including sub-10 nm ones. The multi-resonance optical response of proposed core-shell-like nanostructures (CSLNs) is susceptible to precise tuning by a simple adjustment of one or more of over 10 available fabrication parameters. Notably, already by altering the evaporated metal layer thickness, nanoparticle diameter, interparticle spacing and material selection, the characteristic plasmonic resonances can be placed virtually anywhere within the UV-Vis-NIR spectral range. The dominant spectral resonance is associated with the symmetric nanogap mode which concentrates light in deep subwavelength volumes, yielding enhancement factors in SERS spectroscopy on the order of $10^6$ in practice, but potentially able to reach enhancement factor (*EF)* of $10^9$, in line with the theoretically predicted values for sub-nm nanogaps.

In molecular detection experiments using p-mercaptobenzoic acid (pMBA) and supplementary analytes, we demonstrated that the proposed nanoarchitecture is suitable for comprehensive SERS analysis. We addressed issues related to characterization methodologies, providing insights into the suitability of this architecture for SERS spectroscopy applications, and showed the versatility and robustness of the platform for the purposes of this method. Specifically, it was demonstrated that the CSLN substrates fulfill all five criteria defining a high-quality SERS substrate:[30, 31] uniform signal intensity with variability as low as 2.8%, decent EF of ~$10^6$, sufficient substrate-to-substrate reproducibility (signal variation <15%), high temporal signal stability (signal loss <4%), and documented SERS activity towards three analytes with different charge state. The Langmuir adsorption model was also confirmed for pMBA and the ability to quantitatively detect this analyte using SERS with LOD of $10^{-7}$ M was demonstrated. A thorough investigation of the influence of the architecture parameters of CSLN substrates on the optical response and corresponding SERS signal revealed the capabilities of the platform in a real-life applications, proving the established methodology enables easy, cost-effective, and macroscopic scale fabrication of nanostructured plasmonic substrates with single-nm features. This combination of typically mutually exclusive attributes makes the proposed nanostructures outstanding candidates for other applications requiring a multimodal plasmonic response, high-precision tuning capabilities, and excellent reproducibility.

## Materials and methods

**Substrate preparation:** Microscopic glass slides (Equimed) cut into 2.5 x 2.5 cm pieces were thoroughly rinsed with ethanol (Sigma Aldrich, HPLC, ≥ 99.8% pure) and pure water (0.12 μS/cm), and dried with nitrogen gas (N5.0 purity). Then, they were cleaned in oxygen plasma using the Diener Zepto RIE set to 30% of power for 5 minutes. These two steps were repeated twice. An aqueous solution of poly(diallyldimethylammonium chloride) (PDDA polymer, Merck, 99.5% pure) at a concentration of 0.2 wt% was then pipetted onto the glass and rinsed after 40 s. The same protocol was repeated using an aqueous suspension of sulfate latex beads (composed of sulfate functionalized polystyrene) of the desired diameter and appropriate concentration (ThermoFisher, 8% w/v, diameters: 0.3 μm, 0.2 μm, 0.1 μm and 0.06 μm). Finally, various types of coatings were evaporated onto the substrates using a PVD75 ePVD system (Kurt J. Lesker) with evaporation rates ranging from 0.3 to 1 Å/s. Silver and gold pellets (99.99% pure), as well as germanium pieces (99.999%) were purchased from Kurt J. Lesker. All the processing gasses used for plasma cleaning and evaporation, *i.e.* $O_2$ and $N_2$ were of N5.0 purity.

**Substrate characterization:** UV-Vis spectra of the plasmonic substrates were collected in a transmittance mode using a Metash UV-6100 UV-Vis spectrophotometer and/or a Woollam RC2 ellipsometer (D+NIR model) operating in the wavelength range of 190 nm - 1100 nm and 193 nm - 1690 nm, respectively. SEM images were taken using Zeiss Sigma scanning electron microscope using the InLens detector. The accelerating voltage varied from 8 to 14 kV, depending on the sample and measurement angle.

Nanogap and CSLNs sizes, along with their standard deviations (SD), for metal-coated dielectric nanospheres diameter were determined using ImageJ software (Figure S8 and Table S1). For each SEM image, the contrast was adjusted to most accurately represent the actual size of the nanospheres. The area of each circle was calculated with ImageJ and converted to diameter. The SD was subsequently calculated using Microsoft Excel software.

**FDTD calculations**: Finite-difference time-domain (FDTD) simulations were performed with Ansys Lumerical 3D FDTD commercial software. Optical properties of an amorphous array of CSLNs were simulated with the use of periodic boundary conditions and subwavelength size of the unit cell equal to 150 nm, which corresponds to the average distance between particle centers observed in the experiment. Since the structure exhibits sub-nm details, simulations were carried out with 0.5 nm resolution in the whole simulation region, while in the nanogap additional mesh with size of 0.3 nm was used. Complex refractive indices of metals used in the simulations, namely Ag, Au and Ge, were taken from ellipsometric measurements carried out in the experiment with use of a dual rotating compensator spectroscopic ellipsometer (Woollam RC2). Dielectric spheres (with a refractive index of 1.59 to model latex) were placed on a $SiO_2$ glass substrate with a refractive index taken from Palik.[46] The structure was illuminated from the top (air) side by a broad-band (300 – 1100 nm) plane wave source; transmitted (and reflected) power was collected with a monitor placed below the structure within the substrate. The incident light was polarized in the plane of incidence. Prior to the main simulations, convergence testing of results was performed with use of multiple parameters including simulations time, resolution, number of Perfectly Matched Layers, *etc*.

**Chemicals for the molecular layers formation:** p-mercaptobenzoic acid was purchased from Sigma-Aldrich (Merck), while pyridine and methyl orange were supplied by Ubichem Limited and Thermo Scientific, respectively. KCl and absolute ethanol were purchased from POCh S.A. All the chemicals were used as received without any purification. The aqueous solutions were prepared using ultrapure water with resistivity controlled at 18.2 MΩ·cm, obtained with RephiLe Genie Water system (RephiLe Bioscience, Ltd.).

**Raman instrumentation:** SERS measurements were performed using LabRam HR800 (HORIBA Jobin Yvon) Raman spectrometer, coupled to a BX41 Olympus confocal microscope. All spectra were collected in a back-scattering configuration. A diode-pumped, frequency-doubled Nd:YAG laser operating at a wavelength of 532.0 nm was used as an excitation source. Alternatively, a built-in He–Ne laser providing 632.8 nm line was used. The output power on the sample was below 3 mW for both applied laser lines, focused with a 100x objective lens (Olympus, NA=0.9) or 50x long distance objective (Olympus, NA=0.5). The scattering signal after passing through an edge filter was dispersed using a holographic grating with 600 grooves/mm onto a Peltier-cooled, charge-coupled device (CCD) detector (1024 × 256 pixel) with a working temperature of -70°C. The system was calibrated using the Raman band at 520 cm$^{-1}$ of a silicon wafer, which was also used to control the temporal stability of the Raman set-up performance.

**Sample preparation and data collection in SERS spectroscopy**: SERS activity of the fabricated nanostructures was examined with pMBA as a probe molecule. Functionalization of the surface was carried out by submerging the entire, freshly prepared substrate in pMBA solution in ethanol overnight before SERS studies for each of studied CSLNs, including batch-to-batch comparison. Washing the chemically modified nanostructures with copious amounts of ethanol and drying in air was applied prior to SERS measurements. To evaluate the stability of SERS performance, the as-prepared substrates were first stored in a desiccator for around 4 months and next functionalized with pMBA, according to the procedure described below.

A $10^{-4}$ M pMBA solution was used as a typical concentration for testing the SERS response of the nanostructures prepared under various conditions of synthesis. Further dilution with ethanol down to $10^{-7}$ M was used to identify the limit of detection (LOD) of pMBA by the SERS method for the examined substrates.

SERS spectra of pMBA were typically collected with a 100x lens for each point of the 15×15 grid within a previously defined rectangular range of 61 μm × 58 μm using an automated microscope stage. Three various regions were analyzed for each substrate and particular excitation wavelength with around 1.0 cm lateral spacing from each other. This means that 225 SERS spectra were analyzed for each area, while the SERS signal was measured at 675 various spots for a particular substrate, separated by a macroscopic distance between the three examined areas. Identical conditions were applied for SERS measurements of $10^{-4}$ M of methyl orange adsorbed from aqueous solution.

The SERS measurements of pyridine, used to evaluate the enhancement factor (*EF*), were collected using the same instrument and method as described for pMBA, with two differences. First, a x50 lens was used instead of a x100 lens. Second, rather than using overnight adsorption, the sample was illuminated with a laser through a drop of solution. An appropriate volume of 0.05 M pyridine aqueous solution in 0.1 M KCl was pipetted onto the substrates placed on a glass slide, allowing the solution to flow slightly from the sample onto the slide. The purpose of this approach was to obtain a flatter layer of pyridine solution on the substrate to avoid light lensing from a spherical droplet. The whole procedure was aimed at reproducing the measurement protocol used in the work by Ambroziak *et al.*[68]

The acquisition time for a single accumulation varied from 0.5 to 3 s, while the number of the accumulations was 2 for each spectrum. Single point SERS spectra were acquired typically in a range of 200 – 1800 cm$^{-1}$ and 600 to 1750 cm$^{-1}$ for 532.0 nm and for 632.8 nm excitation lines, respectively.

**SERS data analysis:** The accumulation time varied between the experimental series, but all the values of SERS intensity were recalculated to cts s$^{-1}$ (counts per second), to allow their direct comparison. SERS spectra were first baseline corrected in LabSpec 5 software prior to further analysis. VBA macro executed in Microsoft Excel facilitated the determination of the average (arithmetic mean) maximum intensities of the selected pMBA and methyl orange bands and their relative standard deviation (RSD) within the examined SERS spectra set. RSD were defined as the standard deviation divided by the average SERS intensity in the maximum of a given band, and expressed in percentage and used to evaluate the spatial reproducibility of the SERS signal for a particular CSLN substrate. 3-axis graph visualization and waterfall graphs of line profiles of SERS spectra were plotted using LabSpec 5 software. Fitting of the SERS spectra obtained for pyridine was performed using a house-made Python scripts suite that has already proven itself in similar applications.[78, 79] The LOD for pMBA was determined using a blank sample corresponding to the SERS intensity for the substrate without adsorbed pMBA (denoted as $I_0$). A plot of $I_{SERS}$ *vs* log$_{10}$ from the pMBA concentration was generated, with the line $I_{SERS}=I_0$ parallel to the abscissa. The intersection of this line with a linear fit was projected onto the abscissa yielding log$_{10}$(LOD). The LOD was then determined as 10 raised to this power.

# AUTHOR INFORMATION

**Corresponding Authors**

**Agata Królikowska** - Faculty of Chemistry, University of Warsaw, Pasteura 1, 02-093 Warsaw, Poland, https://orcid.org/0000-0001-7245-2709; Email: akrol@chem.uw.edu.pl

**Piotr Wróbel** - Faculty of Physics, University of Warsaw, Pasteura 5, 02-093 Warsaw, Poland, https://orcid.org/0000-0003-1975-2117; Email: Piotr.Wrobel@fuw.edu.pl

**Authors**

**Mihai C. Suster** - Faculty of Physics, University of Warsaw, Pasteura 5, 02-093 Warsaw, Poland, https://orcid.org/0000-0002-1948-6341

**Aleksandra Szymańska** - Faculty of Physics, University of Warsaw, Pasteura 5, 02-093 Warsaw, Poland, Faculty of Chemistry, University of Warsaw, Pasteura 1, 02-093 Warsaw, Poland, https://orcid.org/0000-0001-5180-914X

**Tomasz J. Antosiewicz** - Faculty of Physics, University of Warsaw, Pasteura 5, 02-093 Warsaw, Poland, https://orcid.org/0000-0003-2535-4174

**Author Contributions**

M.C.S. and A.S. contributed equally. M.C.S. and A.S. optimized the fabrication protocol. M.C.S and A.S. fabricated and characterized the nanostructures with support from P.W., while A.K. and A.S. conducted SERS experiments. P.W. performed FDTD simulations. T.J.A. contributed to the theoretical analysis and data interpretation. P.W. and A.K. supervised the project. The manuscript was written through contributions of all authors. All authors have given approval to the final version of the manuscript.

**Acknowledgement**

P.W. acknowledges financial support from the Excellence Initiative – Research University Programme (IDUB) (project no. BOB-IDUB-622-228/2022). T.J.A. acknowledges support from the National Science Center, Poland, *via* the project 2019/35/B/ST5/02477. A part of this research was supported by the internal funding at the University of Warsaw within the Excellence Initiative –

Research University Programme (New Ideas 2B in Priority Research Area I, project no BOB-IDUB-622-228/2022; PSP 501-D112-20-1004310 project supervised by A.K.: ,Przesuwanie granic powtarzalnej i ultraczułej detekcji antropogenicznych zanieczyszczeń wody – nowa klasa przestrajalnych nanostruktur plazmonowych do kontrolowanych pomiarów za pomocą spektroskopii SERS') that allowed an extensive SERS characterization of the substrates with the use of an azo dye.

The authors would like to acknowledge Robert Ambroziak (Institute of Physical Chemistry, Polish Academy of Sciences) and Marcin Witkowski (Faculty of Chemistry, University of Warsaw) for writing and sharing the scripts for Raman data analysis. The authors also thank the research group of Paweł Majewski (Faculty of Chemistry, Biological and Chemical Research Center, University of Warsaw) for providing access to their chemical lab.

**Supporting Information Available:**

Details of the CSLNs fabrication procedure and visualization of substrate homogeneity at the nanoscale (SEM images) and macroscale (photography); SEM micrographs showing structural details of CSLN geometry; SEM and absorbance results illustrating the effect of Ge wetting layer; FDTD simulations of optical properties for regular CS nanostructures (with fixed geometry or variable metal layer thickness); experimental results of the effect of Au thin film addition on absorbance; typical SERS spectrum of pMBA with vibrational assignment of the bands; determined sizes of the nanogap for CSLNs with different geometries (analysis of SEM images); evaluation of dispersion of diameter size for metal-coated DNSs (based on SEM images); SERS results confirming: sufficient temporal stability; high substrate-to-substrate reproducibility; and SERS activity toward three different analytes.

# Supporting Information for: Nanogap-Engineered Core-Shell-Like Nanostructures for Comprehensive SERS Analysis

Mihai C. Suster†¶, Aleksandra Szymańska‡†¶, Tomasz J. Antosiewicz†, Agata Królikowska‡*, and Piotr Wróbel†∗

*†Faculty of Physics, University of Warsaw, Pasteura 5, 02-093 Warsaw, Poland*
*‡Faculty of Chemistry, University of Warsaw, Pasteura 1, 02-093 Warsaw, Poland*

¶Contributed equally to this work

E-mail: akrol@chem.uw.edu.pl; Piotr.Wrobel@fuw.edu.pl

## S1. FABRICATION PROCEDURE OF CSLNs

Amorphous arrays of Core-Shell-Like Nanostructures (CSLNs) were prepared by the self-assembly of Dielectric NanoSpheres (DNSs) of four distinct diameters ($d$): 60 nm, 100 nm, 200 nm, and 300 nm. Ge films of 1.5 nm thickness and Ag films of various thicknesses (5 nm - 50 nm in height ($h$)) were deposited over the DNSs by the ePVD (electron beam physical vapor deposition) technique. A 5 nm thick layer of Au was also evaporated on selected substrates. The following fabrication process was confirmed to operate on many typical solid supports used in nanomanufacturing, including microscopic soda-lime glass slides and wafers made of silicon, sapphire, GaAs, *etc*. Here, the term 'support' refers to a bare solid substrate, while 'substrate' is used to describe a solid support covered with plasmonic nanostructures.

One of the most challenging aspects of working with colloidal nanoparticles is ensuring their homogeneous distribution across the entire surface of a substrate with no signs of oligomerization. Here, three distinct procedures were applied to address this issue. First, the DNSs coated with negatively charged surface functionalities were selected, which prevents agglomeration of DNSs in suspension. Second, DNSs suspensions of very low concentrations were used to minimize the direct contact between two or more support-attached DNSs. Third, before deposition of the DNSs, the glass supports were coated with a few nanometers thick layer of poly(diallyldimethylammonium chloride) (PDDA). This cationic polymer ensures no migration of the DNSs across the substrate surface due to the electrostatic attraction between the polymer and the DNSs. In addition to these three factors, the development of a reliable cleaning procedure (see below for details) contributed to the spatial uniformity of the CSLNs. The effectiveness of this approach is demonstrated in the SEM images in Figure S1, which display a large-area spatial homogeneity for the metal coated DNSs of two different diameters. The mean distance between the DNSs deposited on the support is concentration-controlled. In addition, the inset in Figure S1a illustrates the macroscopic uniformity of the entire substrate, following the fabrication protocol optimized for Surface-Enhanced Raman Scattering (SERS) analysis.

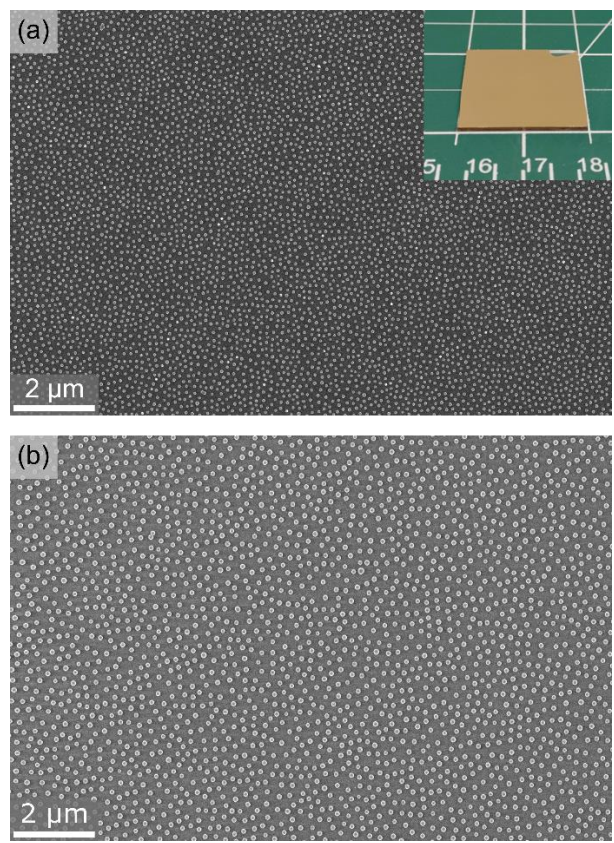

**Figure S1: SEM micrographs confirming high spatial uniformity of CSLN substrates over a large area. The proposed fabrication procedure makes it possible to achieve presented homogeneity across an area of >6 cm$^2$. (a) 60 nm DNSs, (b) 100 nm DNSs, both covered with 1.5 nm Ge and 20 nm Ag. Inset: photograph of a 2.5 cm x 2.5 cm substrate showing the uniformity of the CSLNs composed of 60 nm DNS, coated with 1.5 nm Ge, 40 nm Ag, and 5 nm Au (final geometry optimized for SERS performance). The absence of the metal layer in the upper right corner of the substrate is due to the metal clips holding and masking this area of a sample during the evaporation process. The scale in the inset is marked in centimeters.**

Results presented in this paper were acquired using microscopic glass slides as supports, cut into 2.5 cm x 2.5 cm pieces to enable optical transmission measurements, as well as to reduce the overall manufacturing cost. The cleaning process of the support consisted of two stages: macroscopic and microscopic. The purpose of the first one is to remove any dust, fingerprints, or other relatively large objects from the surface by thoroughly rinsing with ethanol and deionized water. The second stage eliminates further contaminants at the molecular scale by oxygen plasma treatment with the plasma cleaner. This cleaning procedure should be repeated two or more times, depending on the type and condition of the solid support. When performed properly, the resulting cleanliness and wettability allow various liquids to form a uniform film across the entire surface of the glass. In the next step, an adequate amount of 0.2 wt% PDDA solution suspended in water (typically 1 ml) was evenly pipetted over the entire surface of the support, ensuring no overspill. Excess PDDA was then rinsed off with pure water and blown away by a stream of compressed nitrogen or argon gas. After applying such an interlayer, the DNSs suspended in water were pipetted onto the samples in the same manner as the PDDA.

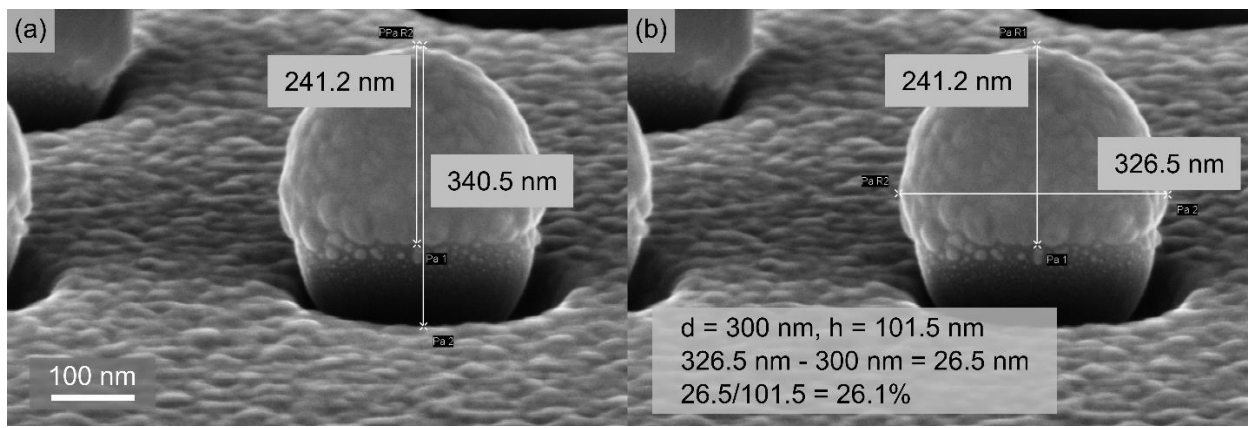

**Figure S2: SEM micrographs presenting the details of CSLN geometry, such as the height and width of the metal cap, as well as the characteristic aperture for a nanostructure consisting of a 300 nm DNS diameter, 1.5 nm Ge and 100 nm Ag layer. The nominal values of a diameter ($d$), and an evaporated metal height ($h$) are given in the legend, together with a calculated shell thickness on the side of the sphere, and a determined shell-to-height ratio.**

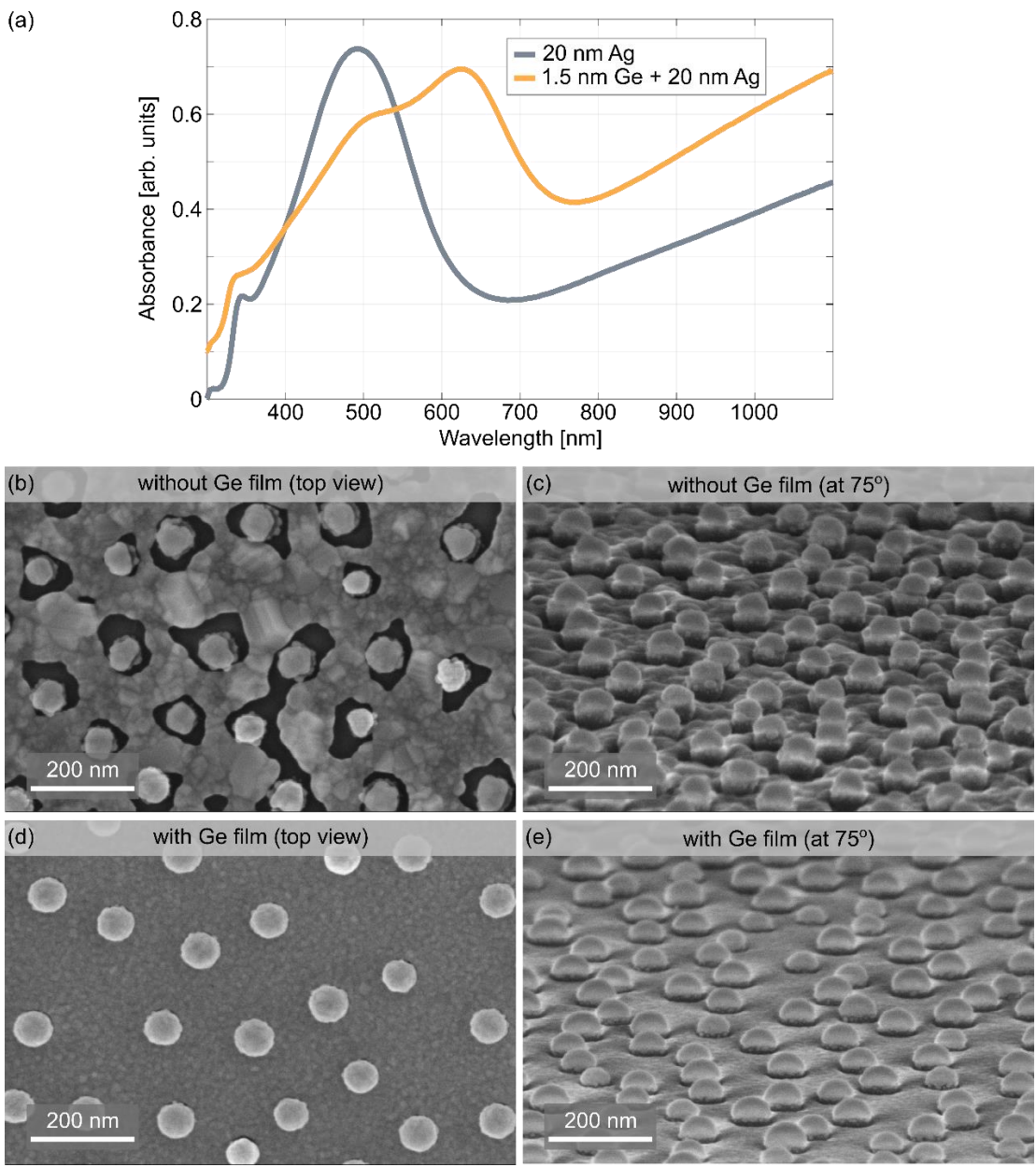

**Figure S3: (a) Absorbance spectra of Ag coated CSLNs with and without a 1.5 nm Ge wetting layer ($d$ = 60 nm, $h$ = 20 nm), (b)-(d) SEM micrographs illustrating the effect of a wetting layer on topography and structural details of the CSLNs. Images (b) and (d) were taken at 90° to the substrate plane, while images (c) and (d) were taken at 75° angle.**

Finally, various types of metal coatings were evaporated onto the DNSs covered supports using a PVD75 Kurt J. Lesker ePVD system, with evaporation rates ranging from 0.3 to 1 Å/s. The total thickness of the evaporated layers and multilayers shown in this paper varied from 6.5 nm to 56.5 nm. Exemplary geometric parameters, such as the height (*h*) and width of the metal cap, are shown in SEM images in Figure S2, presenting a CSLN composed of a 300 nm DNS diameter (*d*) and a 100 nm thick Ag layer. An aperture surrounding the bottom of each nanosphere is also clearly visible. To achieve both uniform and smooth coverage over the DNSs, it is strongly recommended to form a wetting layer prior to evaporation of the plasmonic metal (here: 1.5 nm of Ge in the case of Ag coating). This is important, particularly for thin Ag layers (< 20 nm), as they usually do not form a continuous film when deposited alone.

The effect of a wetting Ge layer versus its absence on the optical response and geometric features of the nanostructures is presented in Figure S3. When an Ag film is deposited directly onto the glass support covered with DNSs (SEM images in Figure S3b and c) it creates a rough semi-continuous coating, with visible large grains. Due to high curvature of the DNS, which further reduces its wettability, Ag does not form a well-adhering cap over the DNS. Instead, it creates a single silver nanoparticle of a poorly (and differently) defined shape on top of each DNS. This is further confirmed by the absorbance spectrum, which for this type of structure resembles the optical response of a silver nanoparticle (Figure S3a; gray curve). However, when a wetting layer is deposited underneath the Ag film (SEM images in Figure S3d and e), the grains become barely visible, the continuity of the layer is greatly improved, and the DNSs metallic coating adheres properly. This improvement is also reflected in the change of absorbance spectrum (Figure S3a; orange curve), which shows three maxima typical of core-shell-like nanostructures, with the highest intensity of the plasmon mode associated with electromagnetic (EM) field enhancement in the nanogap region. This demonstrates that ensuring good wettability of the substrate is essential for this architecture to obtain CSLNs geometry that will generate nanogap mode enhancement.

## S2. CORE-SHELL-LIKE OPTICAL RESPONSE

Figure S4 shows the FDTD calculated scattering, absorption, and extinction cross-sections of fully metal-coated latex-Ag core-shell (CS) nanospheres placed on a semi-infinite glass support. This system, consisting of a latex core sphere with a diameter of 60 nm and an Ag shell thickness of 20 nm, exhibits three distinct spectral features associated with plasmonic resonances present in the extinction cross-section curve.

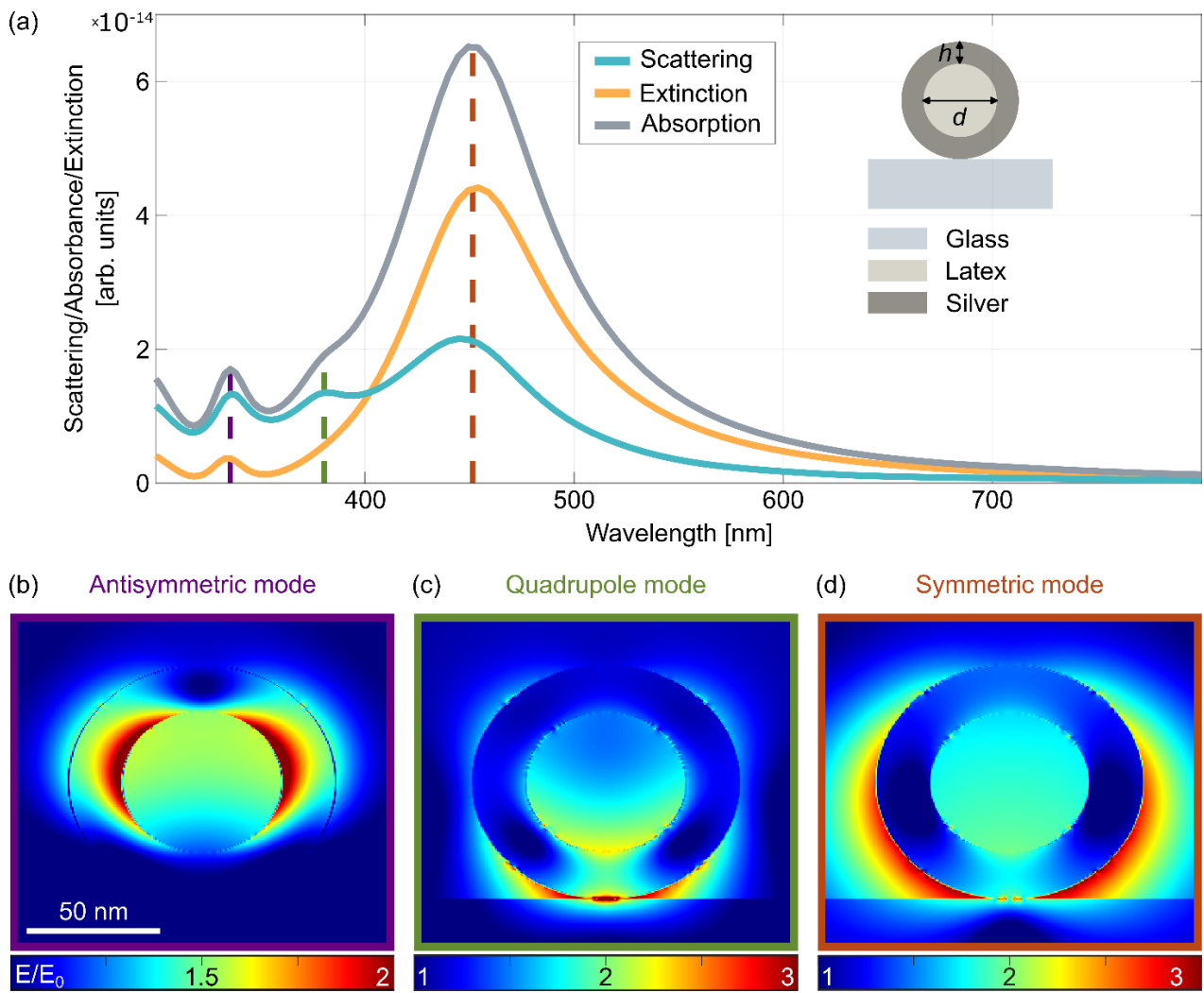

**Figure S4:** (a) FDTD-calculated extinction, scattering, and absorption cross-sections of a fully metal-coated core-shell nanosphere with a 60 nm latex core (modeled with a refractive index of $SiO_2$), and an Ag shell of thickness $h$ = 20 nm on a glass substrate (see the model geometry on the right). (b)-(d) Electric field distributions for the structure depicted in (a), corresponding to the plasmonic modes calculated at the resonant wavelengths indicated by the color-coded dashed lines in panel (a).

The most prominent maximum at about 450 nm corresponds to the symmetric dipolar mode with its corresponding electric field distribution shown in Figure S4d. The maximum at 350 nm is associated with the antisymmetric mode, whose EM field distribution is concentrated at the inner core-shell interface (Figure S4b). The intermediate resonance occurring at 380 nm is attributed to a higher-order quadrupole mode (see Figure S4c for the electric field distribution) and can be observed in case of the large overall size (including both dielectric core and metal shell) of the nanoparticle (*i.e.* 100 nm), while it is absent in the spectra of smaller CS nanoparticles.

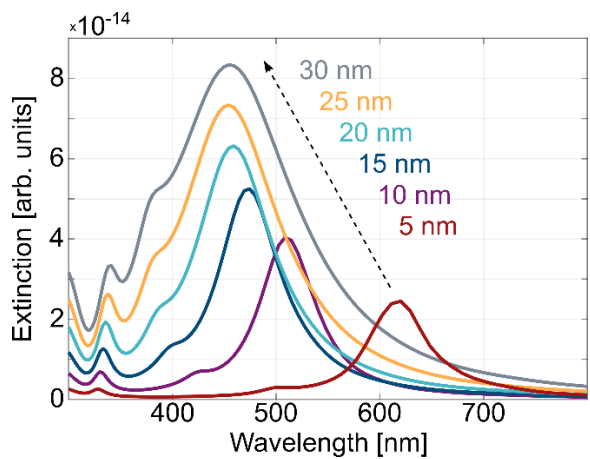

**Figure S5: FDTD-calculated extinction cross-sections of the fully metal-coated core-shell nanosphere, composed of a 60 nm diameter (*d*) latex core and Ag shell thickness (*h*) varying from 5 nm to 30 nm. An arrow indicates a blue shift of the symmetric plasmonic mode with increasing shell thickness.**

The occurrence of the symmetric and antisymmetric modes is characteristic of core-shell nanospheres. Their presence in the extinction spectrum can be distinguished by analyzing their spectral positions as a function of shell thickness (see Figure S5 for the results of FDTD simulations). In the case of a 5 nm thick metal shell, these two resonances are separated by a few hundred nanometers (see a red curve in Figure S5). Increasing the thickness of the metal shell leads to a blue shift of the symmetric mode and a red shift of the antisymmetric mode towards the spectral position corresponding to the dipolar response of a solid metal nanoparticle. The two analyzed resonances do not overlap but remain separated for the thickest simulated metal layer (30 nm in Figure S5; gray curve), while the quadrupole mode resonance appearing in between them becomes more pronounced with the growing thickness of the metal shell. The characteristic multi-resonance optical response (Figure 2a in the main manuscript), along with the electric field distributions at the corresponding resonances (Figure 2b-d) and the dependence of resonance positions on shell thickness in the proposed CSLN geometry (Figure 3a and b), confirm their similarity to fully coated

CS nanoparticles. Additionally, the presence of the nanogap enables control of light in the strongly subwavelength region, surpassing the capabilities of classical CS systems.

**S3. EFFECT OF A THIN GOLD COATING ON PLASMONIC CHARACTERISTICS**

Motivated by the goal to improve the chemical stability and biocompatibility of substrates, we investigated the effect of a thin (5 nm) top layer of Au on the plasmonic properties of CSLNs. Figure S6 highlights the optical differences between the selected substrates with identical fabrication parameters, except for the overall thickness and/or the composition of the metal (multi)layer. The absorbance curves were collected for CSLNs consisting of 60 nm DNS coated with 1.5 nm Ge and 20 nm (gray), 25 nm Ag (cyan) and 20 nm Ag capped by 5 nm Au (orange). Understandably, the general pattern of the absorbance curve changes slightly upon replacement of 5 nm Ag with Au (*cf.* cyan and orange curves). The resemblance of absorbance curves corresponding to CSLNs with a 25 nm thick Ag layer and an Ag/Au bilayer (25 nm in total) indicates that the overall metal thickness is the primary factor governing the shape and position of the plasmonic resonances in the proposed architecture. The introduction of a 5 nm thick gold layer, exhibiting higher extinction coefficients than silver, does not degrade the plasmonic resonances associated with symmetric and antisymmetric modes, preserving the optical quality of the nanostructure while significantly enhancing its chemical stability and durability.

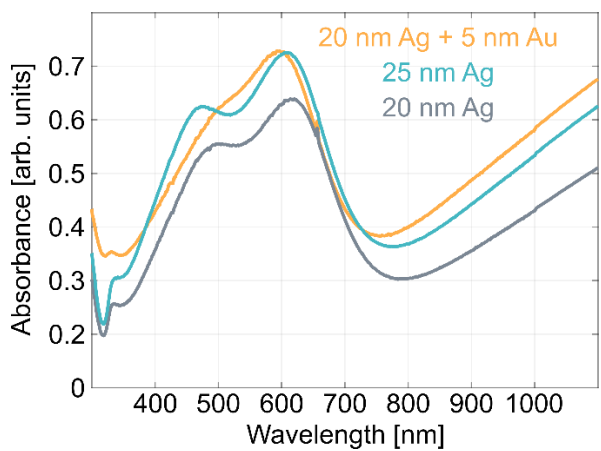

**Figure S6: Effect of a 5 nm thick gold layer on the absorbance spectra of CSLNs formed by 60 nm DNSs (0.2 wt% suspension) coated with 1.5 nm Ge and a 20 nm Ag layer compared to CSLN without Au layer and solely Ag layer of thickness 20 nm and 25 nm.**

## S4. PARAMETERS OF CSLN SUBSTRATES CRITICAL FOR SERS PERFORMANCE

A typical experimental SERS spectrum of p-mercaptobenzoic acid (pMBA) excited with a green laser (532.0 nm) for the molecular layer adsorbed on CSLN substrate fabricated using parameters optimized through the procedure described in the main manuscript, is presented in Figure S7. The molecular formula of pMBA and complete vibrational assignment of its main SERS bands is shown in the inset of Figure S7. For a detailed discussion, the reader is referred to the "SERS performance of CSLNs substrates" section of the manuscript.

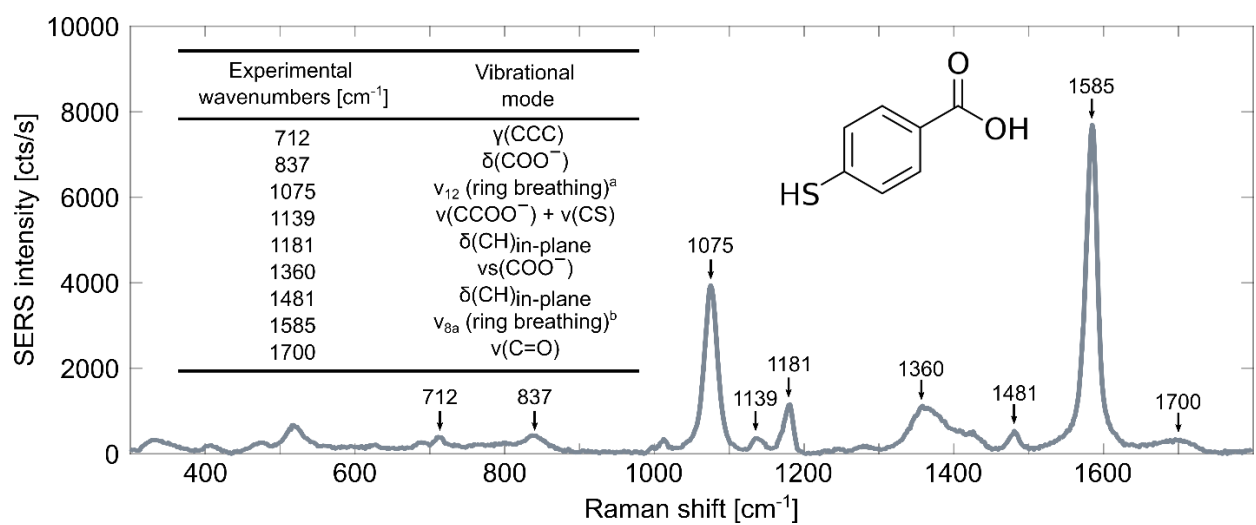

**Figure S7: Typical SERS spectrum (excited with 532.0 nm laser) of pMBA SAMs grown overnight from $10^{-4}$ M ethanolic solution on CSLNs fabricated with 0.2 wt% of 60 nm DNSs, coated with 1.5 nm Ge, 40 nm Ag and 5 nm Au. Inset: molecular formula of pMBA and a table comprising the most intense SERS bands of pMBA. The peak positions and vibrational assignment of major SERS bands observed for the pMBA layer are based on literature reports.[1-4] Greek symbols denote types of vibrations: γ – out-of-plane deformation, δ – bending, ν – stretching. [a and b] Wilson's notation for normal modes of benzene.**

A large part of the analysis in the main manuscript focuses on the effect of plasmonic enhancement through nanogap mode on the SERS signal. To estimate the nanogap size for different DNSs diameters, the structural analysis presented in Figure S8 was performed. SEM images taken at 75° were used to measure the distance from the edge of the aperture to the bottom border of the metal cap coating DNS, considered the most reliable evaluation of the nanogap size. The largest measurement uncertainty is observed for the geometry shown in Figure S8a, which corresponds to the smallest DNSs size and, consequently, the smallest nanogap size. In this case one can also see a slight non-uniformity around the perimeter of a given DNS.

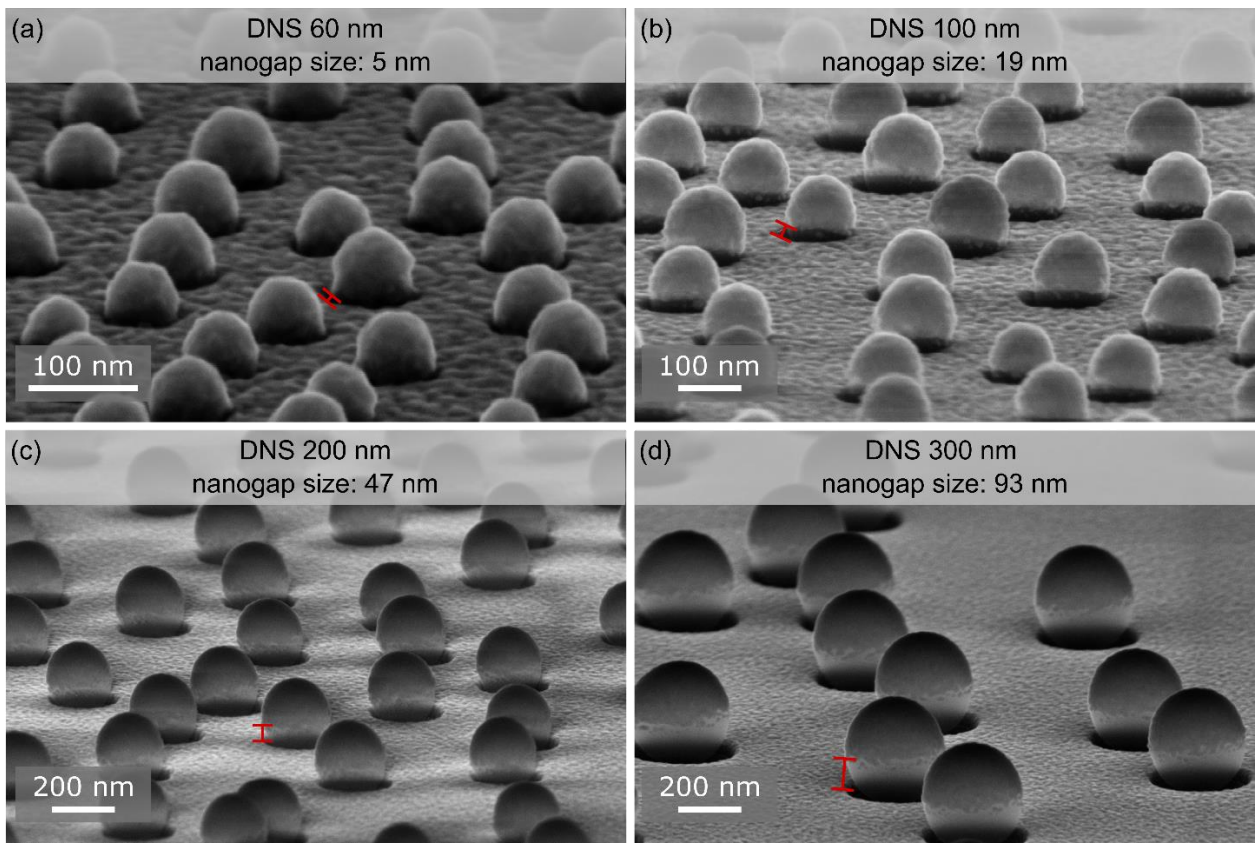

**Figure S8: SEM images comparing nanogap sizes for CSLNs fabricated with DNSs of varying diameters: (a) 60 nm, (b) 100 nm, (c) 200 nm, and (d) 300 nm, each coated with 1.5 nm Ge, 40 nm Ag, and 5 nm Au. One object was analyzed for each case. Red markers indicate typical regions selected for estimating nanogap dimensions. The DNS diameters and corresponding nanogap sizes are given in the legend.**

Statistical analysis of additional SEM images was performed to assess the size dispersion of metal-coated DNSs diameter, which impacts the relative standard deviation (RSD) of the SERS spectra. The diameters of at least 43 DNSs were determined from each SEM image, and the standard deviation (SD) for each substrate was calculated. The results presented in Table S1 align closely with the observed RSD values of the SERS spectra for DNS of varying size and constant total metal layer thickness. Comparison of the data from Table S1 with Figure 6g in the main manuscript reveals a clear relationship between DNS size dispersion and RSD of the SERS signal. The higher the SD of metal-coated DNS diameter, the greater the RSD in SERS signal, resulting from a less uniform geometry of the DNSs comprising an amorphous array of CSLNs and thus its less well-defined plasmonic response.

Figures S9, S10, and S11 prove that CSLN plasmonic substrates meet the criteria #2, #3, and #5 described in Table 1 in the main text (with criteria #1 and #4 addressed in the main manuscript).

Results presented in these figures confirm sufficient temporal stability, high substrate-to-substrate reproducibility, and SERS activity toward three different analytes, respectively.

**Table S1: Effect of DNS size – expressed as DNS diameter ($d$) plus Ag and Au total thickness, determined from the analysis of SEM images – on the resulting SD values for such calculated mean diameters of metal-coated dielectric nanospheres. At least 43 objects were analyzed for each diameter of DNS, evaporated with nominally 40 nm Ag and 5 nm Au thick metal multilayer.**

|  | Nominal diameter of the uncoated DNS | | | |
| --- | --- | --- | --- | --- |
|  | $d = 60$ nm | $d = 100$ nm | $d = 200$ nm | $d = 300$ nm |
| **Mean diameter [nm]** (DNS diameter + Ag and Au thickness) | 73.7 | 117.1 | 207.4 | 293.8 |
| **Standard deviation; SD [nm]** | 5.4 | 7.7 | 7.3 | 24.4 |

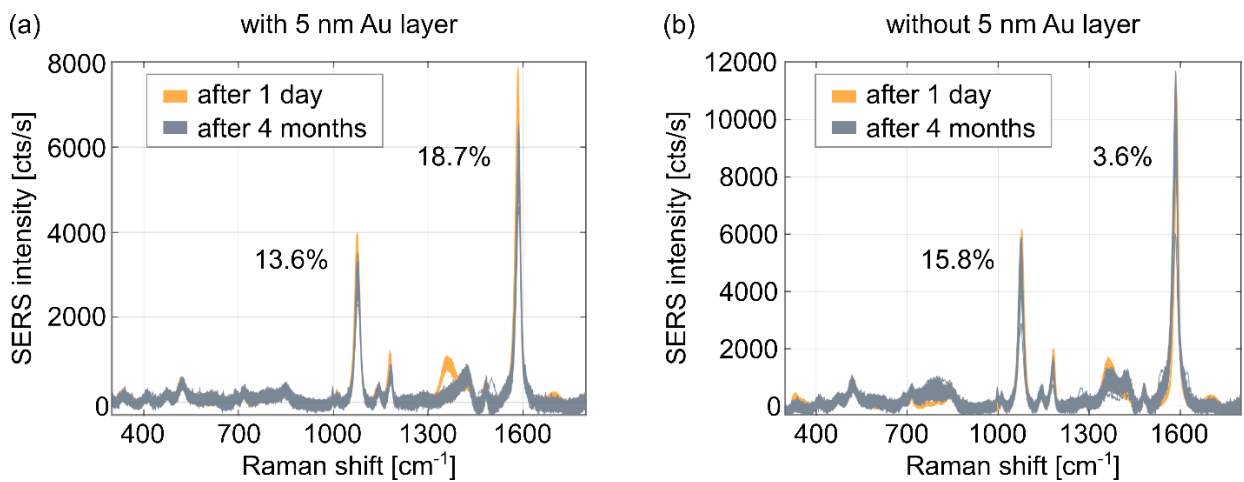

**Figure S9:** Demonstration of the high-quality temporal stability of CSLN substrates: SERS measurements (excited with 532 nm laser) performed on freshly prepared substrates and 4 months after fabrication. The substrates were fabricated with DNS of $d = 60$ nm and $h = 40$ nm of Ag, either (a) with a 5 nm thick Au layer or (b) without the Au layer. The percentage loss of SERS intensity over 4 months for the two strongest bands in the SERS spectrum of $10^{-4}$ M pMBA is indicated in the legend.

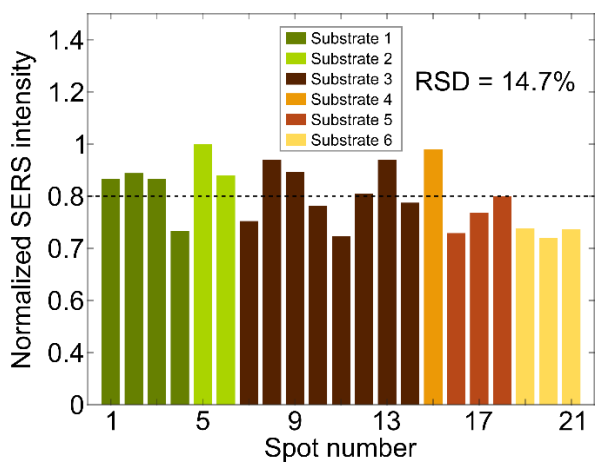

**Figure S10:** Data confirming sufficient substrate-to-substrate SERS reproducibility of CSLNs. Each bar represents an average signal from 225 SERS spectra (excited with 532 nm laser) of $10^{-4}$ M pMBA (band at 1585 cm$^{-1}$) collected for each spot from around 3600 μm$^2$ rectangular area of 6 substrates fabricated over a period of 5 months. The relative standard deviation (RSD) of the signal across all examined 21 spots is 14.7%. The SERS intensities for each bar are normalized to the highest value.

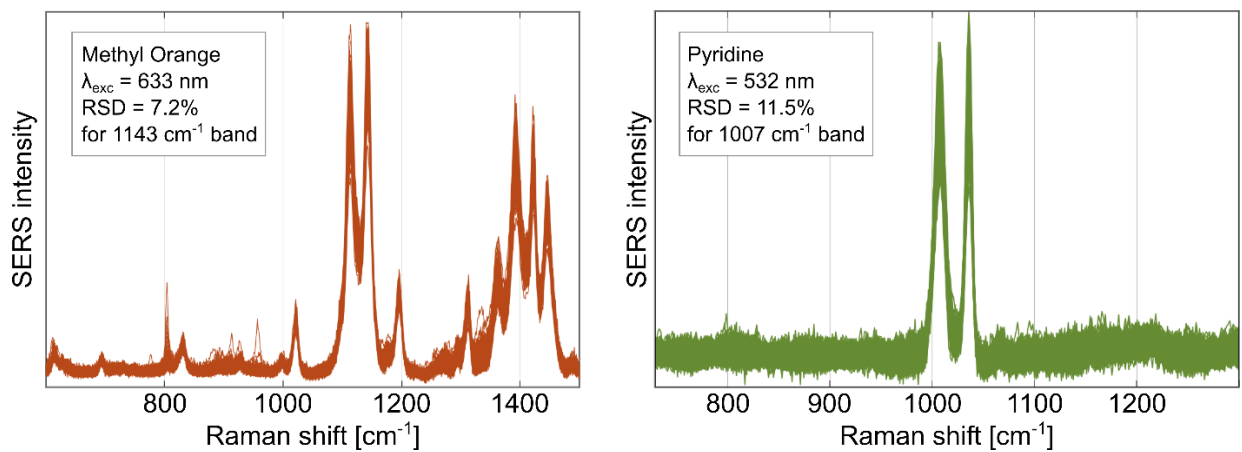

**Figure S11:** 225 SERS spectra of (a) $10^{-4}$ M methyl orange solution (MO) and (b) 0.05 M pyridine (Pyr) in 0.1 M KCl solution, both collected for analytes adsorbed on optimized CSLN substrates with $d$ = 60 nm coated with and $h$ = 40 nm of for Ag and a 5 nm of Au layer. The collection of SERS spectra for Pyr is a part of one of the data sets used to determine the enhancement factor (*EF*). Excitation wavelengths and RSD values of SERS intensity are given in the legend.